\documentclass[a4paper]{article}
\usepackage[normalem]{ulem}
\usepackage{jheppub}
\usepackage{dsfont}
\usepackage{amsmath,amssymb,amscd,amsfonts,mathtools}
\DeclareMathOperator{\Tr}{Tr}

\usepackage[toc,page]{appendix}
\usepackage{epsfig}
\usepackage{epstopdf}
\usepackage{latexsym}
\usepackage{graphicx}
\usepackage{placeins}
\usepackage{floatrow}
\usepackage{subfig}
\usepackage{caption}
\usepackage{booktabs}
\usepackage{bbm}
\usepackage{color}
\usepackage{physics}
\usepackage{tensor}
\usepackage{tikz}
\usetikzlibrary{matrix}
\usetikzlibrary{decorations.markings,calc,shapes,decorations.pathmorphing,patterns,decorations.pathreplacing}
\usetikzlibrary{positioning}
\usepackage{hyperref}

\pdfoutput=1
\makeatletter
\def\@fpheader{\relax}
\makeatother

\title{Anisotropic Flows into Black Holes}

\author{Elena Caceres,}
\author{Sanjit Shashi}
\affiliation{Weinberg Theory Group, Department of Physics, University of Texas,\\
\phantom{}\hspace{0.5cm} 2515 Speedway, Austin, Texas 78712, USA.}
\emailAdd{elenac@utexas.edu}
\emailAdd{sshashi@utexas.edu}

\abstract{We consider anisotropic black holes in the context of holographic renormalization group (RG) flows. We construct an $a$-function that is stationary at the boundary and the horizon and prove that it is also monotonic in both the exterior and the interior of the black hole. In spite of the reduced symmetry, we find that the ``radial" null energy condition is sufficient to ensure the existence of this monotonic $a$-function. After constructing the $a$-function, we explore a holographic anisotropic $p$-wave superfluid state as a concrete example and numerical testing grounds. In doing so, we find that the $a$-function exhibits nontrivial oscillations in the trans-IR regime while preserving monotonicity. We find evidence that such oscillations appear to drive the trans-IR flow into nontrivial fixed points. We conclude by briefly discussing how our work fits into both the broader program of holographic RG flow and quantum information approaches to probing the black hole interior.}

\begin{document}
\maketitle
\flushbottom

\section{Introduction} 
The AdS/CFT correspondence \cite{Maldacena:1997re} provides a framework to study the renormalization group (RG) structure of strongly coupled field theories. In the context of holographic RG flows \cite{Balasubramanian:1999jd}, the radial coordinate
in the bulk corresponds to an energy scale at the boundary and parametrizes the RG flow of the dual theory. Thus, in Einstein gravity coupled to matter, the evolution of bulk fields along the radial direction encodes how observables in the dual field theory change with the energy scale \cite{deBoer:1999tgo,deBoer:2000cz,Bianchi:2001kw,Fukuma:2002sb,Papadimitriou:2004ap,Papadimitriou:2005ii}. Specifically, one considers a relevant operator that deforms the ultraviolet (UV) boundary CFT and triggers an RG flow. By the rules of AdS/CFT, this operator is dual to bulk matter, and the backreaction sourced by that matter describes the flow. 
 
One may immediately ask how to understand black holes in this framework. Holographic RG flow in the exterior region of a black hole geometry can be understood as the RG flow of an excited state between a UV fixed point at the boundary and an IR fixed point at the horizon. In the interior, the situation is more subtle. The radial extra dimension is timelike, and so interior slices describe an imaginary energy domain of the flow. This can be seen as a ``trans-IR" analytic continuation of the ``conventional" UV $\to$ IR part of the flow described by the exterior \cite{Frenkel:2020ysx,Caceres:2022smh}. In other words, the notion of trans-IR flows provides an overarching framework in field-theoretic language with which to conceptualize the black hole interior and its quantum informatics \cite{Caceres:2022smh}.

In \cite{Caceres:2022smh}, we explored whether such holographic trans-IR flows represent coarse graining like conventional flows do. This idea is made concrete by the existence of monotonic functions which decrease from the UV to the IR and thus ``count" degrees of freedom, like Zamolodchikov's $c$-function \cite{Zamolodchikov:1986gt}, Cardy's $a$-function \cite{Cardy:1988cwa,Komargodski:2011vj}, and their holographic analogs \cite{Freedman:1999gp,Myers:2010xs,Myers:2010tj}. We constructed \cite{Caceres:2022smh} a thermal analog to the usual holographic $a$-function of \cite{Freedman:1999gp,Myers:2010xs,Myers:2010tj} which, assuming the \textit{null energy condition (NEC)}, monotonically decreases along the full flow. Thus, we interpreted the trans-IR as describing coarse graining.

However, the black hole solutions we studied in \cite{Caceres:2022smh} were highly symmetric---namely planar and isotropic. Many holographic RG flows, such as those employed in AdS/CMT as toy models of realistic non-Lorentzian systems, are not as symmetric. So, it is natural to ask to what extent the results of \cite{Caceres:2022smh} rely on the symmetric nature of the background. Can we construct a monotonic function along the \textit{entire} flow---including the trans-IR---in less symmetric cases? If so, is the NEC still a sufficient condition? In this work, we answer those questions in the affirmative.

Specifically, we study anisotropic black holes which spontaneously break Lorentz symmetry to investigate whether assuming the NEC is sufficient to construct an $a$-function which is monotonic even in the trans-IR and has fixed points at both the UV and the IR. More specifically, we use only the ``radial" NEC---the condition corresponding to the null vector pointing along the radial extra dimension $\rho$---to construct our $a$-function. From the holographic RG flow perspective, this is the natural thing to do because $\rho$ corresponds to the energy scale parameterizing the flow; this radial NEC essentially imposes positivity of energy (density) along the flow.

A priori, the existence of an $a$-function is not obvious. Indeed, it is known that entanglement entropy, which plays the role of a monotonic function in Lorentz-invariant flows \cite{Casini:2006es,Casini:2012ei}, does not do so in generic Lorentz-breaking flows \cite{Swingle:2013zla}. However, previous work on anisotropic deformations of black holes \cite{Giataganas:2017koz} and vacuum states \cite{Chu:2019uoh,Ghasemi:2019xrl} observed that the reduced symmetry actually gives rise to multiple independent NECs, and by assuming all of them it is possible to construct multiple independent functions (including our own\footnote{In \cite{Giataganas:2017koz}, the radial NEC appears to have a privileged role as being directly derivable from Einstein's equations. So even with multiple independent monotonic functions available, our $a$-function may be said to be more fundamental in that the NEC need not be separately assumed on top of considering Einstein gravity. We thank Juan Pedraza for emphasizing this point.}) monotonic from the UV to the IR. Nevertheless, how these functions behave in the trans-IR is unclear---the focus of all previous work on anisotropic RG flows has been the exterior of the black hole. In studying monotonicity in the trans-IR, we employ just the radial NEC because that is the condition which is most naturally interpreted as energy positivity along the RG flow.

\subsection{Anisotropic Flows}

Concretely, we consider a class of anisotropic flows for which the $(d+1)$-dimensional bulk is Einsteinian and has a geometry of the form
\begin{equation}
ds^2 = e^{2A(\rho)}\left[-f(\rho)^2 dt^2 + e^{2\mathcal{X}(\rho)} d\vec{x}_1^2 + d\vec{x}_2^2\right] + d\rho^2,\label{met}
\end{equation}
where $t \in \mathbb{R}$, $\rho \geq 0$, $\vec{x}_1 \in \mathbb{R}^{\delta_1}$, and $\vec{x}_2 \in \mathbb{R}^{\delta_2}$. $\delta_1$ and $\delta_2$ are integers such that $\delta_1 + \delta_2 = d-1$.

By introducing the warp factor $e^{2\mathcal{X}(\rho)}$, we break the rotational symmetry of the constant-$\rho$ slices as
\begin{equation}
\text{SO}(d-1) \to \text{SO}(\delta_1) \times \text{SO}(\delta_2).\label{symmBreak}
\end{equation}
This is a rather mild breaking of the original rotational symmetry. Thus, although \eqref{met} has reduced symmetry, it should be a tractable setting in which to probe our central questions. Ultimately, in terms of the coordinates of \eqref{met}, we propose that
\begin{equation}\label{eq:our_afunction}
a_T^{(\delta_1)}(\rho) \sim e^{-\delta_1 \mathcal{X}(\rho)} \left[\frac{(d-1) f(\rho)}{\delta_1 \big(A'(\rho) + \mathcal{X}'(\rho)\big) + \delta_2 A'(\rho)}\right]^{d-1}
\end{equation}
is the monotonic $a$-function for flows \eqref{met} whose dynamics are dictated by the Einstein equations. Note that this is also one of the monotones of \cite{Giataganas:2017koz}.

Anisotropic flows of this kind have been studied in the holographic literature as realizations of holographic $p$-wave superfluid states \cite{Gubser:2008wv,Ammon:2009xh}. In those cases, the anisotropy is introduced by a normalizable vector mode in the near-UV which ``picks out" and warps one of the spatial directions, thereby spontaneously breaking full rotational symmetry.\footnote{That this symmetry breaking is spontaneous is because the vector mode decays to $0$ at the conformal boundary. There are also hairy solutions which have a source on the boundary, but we do not consider these in the context of holographic RG flow.} After providing more general arguments about monotonicity, we study a holographic $p$-wave superfluid state as a concrete example.

\subsection{Outline}

In Section \ref{sec:symmetric}, we first review and discuss the constructions of holographic $a$-functions done in symmetric domain-wall solutions, with some technical details discussed in Appendix \ref{app:isoCorrect}. We use the symmetric cases to motivate the central conjecture of this paper: that only the radial NEC corresponding to the direction of the RG flow is needed to guarantee a monotonic $a$-function.

In Section \ref{sec:aniso}, we present a systematic procedure to determine a candidate $a$-function from the radial NEC that is also applicable to the anisotropic ansatz \eqref{met}. This candidate $a$-function supports both the previous cases considered in the literature (vacuum deformations and thermal deformations) and their anisotropic counterparts. However, our initial procedure for construction only guarantees monotonicity of the holographic $a$-function in the exterior. Thus, we prove trans-IR monotonicity separately; this requires changing to a more convenient set of coordinates.

After our general construction, in Section \ref{sec:pwave} we consider the holographic $p$-wave superfluid state constructed by \cite{Ammon:2009xh}. These satisfy the radial NEC, as we show in Appendix \ref{app:necpwave}, and so they support monotonic $a$-functions. We also elaborate on some technical details about the numerical construction of such states in Appendix \ref{app:numerics}. We numerically compute the anisotropic $a$-function and show that it exhibits nontrivial fluctuations in the interior. Interestingly, we find that the derivative of this $a$-function has roots, providing evidence that the fluctuations in $a$ drive this trans-IR flow into novel fixed points. Nonetheless, in spite of the fluctuations, the $a$-function in this holographic $p$-wave superfluid state still exhibits monotonicity.

We conclude with some discussion in Section \ref{sec:disc}. We make some statements about the remaining mysteries surrounding field-theoretic interpretations of trans-IR flows. We also mention how our work may connect to the quantum-information approach to black hole interiors.

\section{Lessons from Symmetric Flows}\label{sec:symmetric}

We first briefly summarize the standard constructions of holographic $a$-functions in the symmetric flows considered in the literature. The arguments presented in this review are sparse, with the details concerning vacuum deformations in \cite{Myers:2010xs,Myers:2010tj} and the details concerning thermal deformations in \cite{Caceres:2022smh}.

These symmetric flows typically only use one of the null energy conditions---the one corresponding to the radial extra dimension of the AdS bulk. Such a NEC is essentially a statement about the positivity of energy density along the direction of the RG flow. So, we conjecture that this radial NEC is all that is needed to guarantee a monotonic $a$-function.

\subsection{Review of $a$-Functions in Symmetric Flows}

Holographic $a$-functions interpolate from the UV trace anomaly coefficient $\ell^{d-1}$ \cite{Henningson:1998gx} (where $\ell$ is the bulk AdS curvature radius). They were first constructed \cite{Freedman:1999gp,Myers:2010xs,Myers:2010tj} in backreacting $(d+1)$-dimensional AdS describing deformations of the UV vacuum state,
\begin{equation}
ds^2 = e^{2A(\rho)}\left(-dt^2 + d\vec{x}^2\right) + d\rho^2,\ \ (t,\vec{x}) \in \mathbb{R}^d,\ \ \rho \geq 0.\label{vacmet}
\end{equation}
In Einstein gravity,\footnote{Note that \cite{Myers:2010xs,Myers:2010tj} also considers Gauss-Bonnet gravity, in which the $a$-function takes a different form. However, the usual argument of monotonicity is the same as in Einstein gravity.} one may suggest that the $a$-function for vacuum deformations takes the form\footnote{There is technically an overall prefactor of $\pi^{d/2}/[\Gamma(d/2)\ell_{\text{P}}^{d-1}]$. Because only the $\rho$-dependence is important, we omit such prefactors involving $d$ and $\ell_{\text{P}}$ for convenience, using $\sim$ in place of $=$ whenever we do so.}
\begin{equation}
a(\rho) \sim \left[\frac{1}{A'(\rho)}\right]^{d-1}.\label{aVac}
\end{equation}
The metric must be asymptotically AdS, so at large $\rho$ we have that $A(\rho) \sim \dfrac{\rho}{\ell}$. By plugging this into \eqref{aVac}, we quickly see that the $a$-function reduces to the holographic trace anomaly in the UV. Once \eqref{aVac} is in hand as a suitable candidate, it is acceptable to set $\ell = 1$ for convenience.

Now, let $k^\mu$ be any null vector in the spacetime of interest. The null energy condition (NEC),
\begin{equation}\label{eq:nec}
T_{\mu\nu} k^\mu k^\nu \geq 0,\end{equation}
imposes constraints on the matter sector of Einstein's equations,
\begin{equation}
G_{\mu\nu} - \dfrac{d(d-1)}{2} g_{\mu\nu} = T_{\mu\nu}.
\end{equation}
Using the NEC corresponding to the radial null vector $k^\mu = e^{-A(\rho)} \delta^\mu_t + \delta^\mu_\rho$ in \eqref{eq:nec}, one can show that \eqref{aVac} is monotonic \cite{Myers:2010xs,Myers:2010tj},\footnote{The proof of monotonicity for odd $d$ is rather subtle. If $A'$ is nonpositive anywhere along the flow, then monotonicity of $a$ is no longer assured. \cite{Myers:2010tj} uses a proof by contradiction demonstrating that the NEC is enough to keep $A'(\rho) \geq 0$, but they only briefly touch on a particular edge case. We provide a more complete proof in Appendix \ref{app:isoCorrect}.}
\begin{equation}
\frac{da}{d\rho} \sim \frac{1}{A'(\rho)^d}\left(\tensor{T}{^\rho_\rho}-\tensor{T}{^t_t}\right) \geq 0,
\end{equation}
validating its interpretation as a count of degrees of freedom along the flow.

To explore monotonicity in the trans-IR, we had constructed a holographic $a$-function from backreacting $(d+1)$-dimensional AdS-Schwarzschild describing deformations of the UV thermal state,
\begin{equation}
ds^2 = e^{2A(\rho)}\left[-f(\rho)^2 dt^2 + d\vec{x}^2\right] + d\rho^2.\label{thermalDomainWall}
\end{equation}
$f$ has a simple root at $\rho = 0$ (the horizon), and the interior is accessed by analytic continuation of $t$ and $\rho$ to complex values,
\begin{equation}
t = t_I - \text{sgn}(t_I) \frac{i\gamma}{2\mathcal{T}},\ \ \rho = i\kappa.\label{analyticContCoords}
\end{equation}
Here, $t_I \in \mathbb{R}$, $\gamma$ is some half-integer, $\mathcal{T} = \dfrac{e^{A(0)}f'(0)}{2\pi}$ is the temperature of the black hole, and $\kappa > 0$. The proposed $a$-function in this geometry is written as $a_T$ and is given as \cite{Caceres:2022smh}
\begin{equation}
a_T(\rho) \sim \left[\frac{f(\rho)}{A'(\rho)}\right]^{d-1}.\label{thermalA}
\end{equation}
To show monotonicity along both the interior and the exterior,
\begin{equation}
\left.\frac{da_T}{d\rho}\right|_{\rho > 0} > 0,\ \ \left.\frac{da_T}{d\kappa}\right|_{\kappa > 0} < 0,
\end{equation}
we may transform $\rho$ to a radial coordinate $\bar{r}$ which is real everywhere, as done in \cite{Caceres:2022smh}. This is done to avoid branch-cut ambiguities arising from imaginary factors arising from analytic continuation of $\rho$ to the interior, but the trade-off is that $\bar{r}$ itself is not a well-defined energy scale since it will need to furnish a coordinate singularity. Specifically, we set
\begin{equation}
e^{2A(\rho)} = \frac{1}{\bar{r}^2},\ \ f(\rho)^2 = F(\bar{r})e^{-\chi(\bar{r})},\ \ d\rho = -\frac{d\bar{r}}{\bar{r}\sqrt{F(\bar{r})}},\label{coordChange}
\end{equation}
where $F$ has a simple root at some $\bar{r} = \bar{r}_h > 0$. This transformation yields
\begin{equation}
ds^2 = \frac{1}{\bar{r}^2}\left[-F(\bar{r}) e^{-\chi(\bar{r})} dt^2 + \frac{d\bar{r}^2}{F(\bar{r})} + d\vec{x}^2\right],\ \ a_T(\bar{r}) \sim e^{-(d-1)\chi(\bar{r})/2}.\label{isotropic}
\end{equation}
Again, monotonicity follows from the NEC of just one radial null vector. Thus, it is meaningful to think of the trans-IR flow as a coarse-graining procedure.

\subsection{Physicality of the Radial NEC} \label{sec:radNEC}

The null energy condition is sufficient, but not necessary, to prove monotonicity. One may ask whether our assumptions may be made laxer. For example, we may consider weaker energy conditions for the matter, such as the average null energy condition (ANEC) \cite{Friedman:1993ty} or quantum null energy condition (QNEC) \cite{Bousso:2015mna,Bousso:2015wca}. From the standpoint of covariance, this seems sensible. However, there is evidence in top-down models that completely sacrificing the NEC also does away with monotonicity of all known holographic $a$-functions \cite{Hoyos:2021vhl}.\footnote{A possible counterpoint here is that the usual holographic $a$-functions are all constructed via the NEC anyway, and so eliminating it would also eliminate such constructions. One convincing argument for this position would be to find monotonic functions which cannot be constructed from the NEC, but rather from some other looser conditions, and which retain monotonicity even when the NEC is violated.} And so, the correct general weakened condition might be less constraining than the full NEC but more constraining than the lack of the NEC.

That being said, in the language of holographic RG flow, it would actually make sense to treat the $\rho$ coordinate as special. This is because the $\rho$ coordinate describes the energy scale. Thus, if we want to describe how degrees of freedom may change along the flow, it is reasonable to constrain \textit{only} a measure of bulk energy density specifically along the flow (i.e. along the $\rho$ direction). To this effect, we conjecture:

\begin{quote}
\textit{Only the radial NEC computed from a null vector pointing along the radial extra dimension is needed to prove the existence of a monotonic function in a holographic RG flow.}
\end{quote}

This is true in the examples above. However, in the case of vacuum deformations \eqref{vacmet}, the conjecture is trivial---any null vector yields the same condition up to multiplication by a nonnegative function.\footnote{This can be shown by explicit computation by contracting the stress tensor $T_{\mu\nu}$ against a generic null vector.} Meanwhile, the statement is not as trivial in thermal deformations \eqref{thermalDomainWall} in which have multiple truly independent NECs depending on the null vector chosen. In that case, the radial NEC is all we need \cite{Caceres:2022smh}. 

Anisotropic flows present another test of our conjecture. Even anisotropic flows from the vacuum, let alone the flows into black holes, support different independent NECs. Previous work \cite{Giataganas:2017koz,Chu:2019uoh,Ghasemi:2019xrl} assumed all of these NECs to construct \textit{multiple} independent monotones. However, we only care about whether one such monotone can be constructed from just the radial NEC and whether the resulting function is monotonic in the trans-IR. We find this to be the case.

\section{Monotonicity in Anisotropic Flows} \label{sec:aniso}

Consider the following $(d+1)$-dimensional domain-wall ansatz,
\begin{equation}
ds^2 = e^{2A(\rho)} \left[-f(\rho)^2 dt^2 + e^{2\mathcal{X}(\rho)} d\vec{x}_1^2 + d\vec{x}_2^2\right] + d\rho^2,\ \ t \in \mathbb{R},\ \ \rho \geq 0,\ \ \vec{x}_1 \in \mathbb{R}^{\delta_1},\ \ \vec{x}_2 \in \mathbb{R}^{\delta_2}.\label{anisoDomainWall}
\end{equation}
Note that $\delta_1 + \delta_2 = d-1$. For now, we place no restrictions on the metric aside from having the constant-$\rho$ slices be Lorentzian on $\rho > 0$, i.e.
\begin{equation}
\rho > 0 \implies f(\rho) > 0,
\end{equation}
and having it be asymptotically AdS with curvature radius $\ell$, which means that at large $\rho$,
\begin{equation}
A(\rho) \sim \frac{\rho}{\ell}, \ \ \mathcal{X}(\rho) \sim 0,\ \ f(\rho) \sim 1.\label{boundaryConds}
\end{equation}
This ansatz not only includes the previous cases of isotropic vacuum and thermal deformations, but it also supports anisotropic deformations of either kind. As such, it is a more generalized domain-wall ansatz than \eqref{vacmet} and \eqref{thermalDomainWall}.

In this section we construct the $a$-function,
\begin{equation}\label{eq:our_afunction2}
a_T^{(\delta_1)}(\rho) \sim e^{-\delta_1 \mathcal{X}(\rho)} \left[\frac{(d-1) f(\rho)}{\delta_1 \big(A'(\rho) + \mathcal{X}'(\rho)\big) + \delta_2 A'(\rho)}\right]^{d-1},
\end{equation}
and prove that it is monotonic both in the exterior and the interior of the black hole. 

Our strategy is to first focus on the exterior of the black hole and arrive at \eqref{eq:our_afunction2} by integrating the radial NEC. This is a more systematic approach than starting with the holographic trace anomaly, and it automatically yields a monotonic $a$-function for anisotropic flows from the vacuum. However, as we are interested in the black hole interior, we subsequently change coordinates to prove that our construction is also monotonic in the trans-IR.

\subsection{$a$-Functions from Integrating the Radial NEC}\label{sec:intNEC}

We first attempt to establish a systematic approach to constructing monotonic $a$-functions starting from the radial NEC. We use the previously established symmetric cases discussed in Section \ref{sec:symmetric} to guide our general approach, but we find that it needs to be generalized slightly to work for anisotropic flows.

First, observe that we can arrive at \eqref{aVac} and \eqref{thermalA} by \textit{starting} with the NEC along some null vector in the $(t,\rho)$ plane. For example, for vacuum deformations and using the null vector $k^\mu = e^{-A(\rho)}\delta^\mu_t + \delta^\mu_\rho$ in \eqref{vacmet}, the NEC reads
\begin{equation}
-(d-1)A''(\rho)\ge 0.\label{radNecVac}
\end{equation}
It helps to rewrite this in the form,
\begin{equation}\label{eq:necSchemeK0} 
\mathcal{C}(\rho) \frac{d}{d\rho}\left[\mathfrak{a}(\rho)^{1/(d-1)}\right] \ge 0,
\end{equation}
where $\mathcal{C}$ is a positive-definite real function while $\mathfrak{a}^{1/(d-1)}$ is the principle branch of the $(d-1)$th root of some real function $\mathfrak{a}$. If we can write the NEC like this, then we may immediately identify a function $\mathfrak{a}^{1/(d-1)}$ which is manifestly monotonic in $\rho$, since its derivative would be nonnegative. For example, to recover \eqref{radNecVac} specifically, we may substitute
\begin{equation}
\mathfrak{a}(\rho) = \left[\frac{1}{A'(\rho)}\right]^{d-1},\ \ \mathcal{C}(\rho) = (d-1)A'(\rho)^2.\label{vacSchemea}
\end{equation}
We identify $\mathfrak{a}$ as our candidate $a$-function \eqref{aVac}. Since $ \mathcal{C} > 0$, the monotonicity of $\mathfrak{a}^{1/(d-1)}$ follows from \eqref{eq:necSchemeK0}. Furthermore, by assuming an asymptotically-AdS metric, the radial NEC may be used to show that, for any $d$,
\begin{equation}
\mathfrak{a}^{(d-2)/(d-1)} = \frac{1}{A'(\rho)^{d-2}} \geq 0,
\end{equation}
or, equivalently, that $A'(\rho)^d \geq 0$---see \cite{Myers:2010tj,Myers:2010xs} and Appendix \ref{app:isoVac}. Thus, because we may write
\begin{equation}
	\frac{d}{d\rho}\left[\mathfrak{a}(\rho)\right] = \frac{d}{d\rho}\left[\left(\mathfrak{a}(\rho)^{1/(d-1)}\right)^{d-1}\right] = (d-1) \mathfrak{a}(\rho)^{(d-2)/(d-1)} \frac{d}{d\rho}\left[\mathfrak{a}(\rho)^{1/(d-1)}\right],
\end{equation}
it follows that $\mathfrak{a}$ is also monotonic, and from \eqref{boundaryConds} the $\mathfrak{a}$ in \eqref{vacSchemea} asymptotes to the appropriate UV trace anomaly coefficient $\ell^{d-1}$.

Similar logic holds for the exterior ($\rho > 0$) region of thermal deformations \eqref{thermalDomainWall}. In this case, we find that the NEC from $k^\mu = \dfrac{e^{-A(\rho)}}{f(\rho)} \delta^\mu_t + \delta^\mu_\rho$ is
\begin{equation}
 \frac{(d-1)}{f(\rho)}\left[f'(\rho)A'(\rho) - f(\rho)A''(\rho)\right] \ge 0.\label{radNecIsoTherm}
\end{equation}
We may write this in the form \eqref{eq:necSchemeK0} by taking
\begin{equation}
\mathfrak{a}(\rho) = \left[\frac{f(\rho)}{A'(\rho)}\right]^{d-1},\ \ \mathcal{C}(\rho) = \frac{(d-1)A'(\rho)^2}{f(\rho)}.\label{thermSchemea}
\end{equation}
We thus recover \eqref{thermalA}, and similar reasoning as in the vacuum case utilizing the asymptotics \eqref{boundaryConds} and the radial NEC implies that this $\mathfrak{a}$ is monotonic in the black hole exterior---see Appendix \ref{app:isoTherm}. Furthermore, observe that this $\mathfrak{a}$ also asymptotes to the UV trace anomaly coefficient. As a good sanity check on our equations, we note that \eqref{thermSchemea} reduces to \eqref{vacSchemea} by forcibly setting $f(\rho) = 1$.\footnote{We point out that one may set $f(\rho) = f_*$, where $f_*$ is any constant, and simultaneously recover the ansatz \eqref{vacmet} (after rescaling the $t$ coordinate) but not obtain the appropriate normalization for $\mathfrak{a}$. However, this choice for $f$ violates the specified boundary condition \eqref{boundaryConds}, so we may only consider the case of $f_* = 1$.}

For anisotropic backgrounds of the form \eqref{anisoDomainWall}, we find it unfeasible to write the radial NEC in the form \eqref{eq:necSchemeK0}. One way forward is to generalize our strategy. Since all we need to do is to ensure the existence of \textit{some} $\mathfrak{a}$ and $\mathcal{C}$ satisfying \eqref{eq:necSchemeK0}, it is sufficient to attempt to write the NEC as
\begin{equation}
\mathcal{C}(\rho) \frac{d}{d\rho}\left[\mathfrak{a}(\rho)^{1/(d-1)}\right] - \mathcal{K}(\rho)^2 \ge 0,\label{necScheme}
\end{equation}
where $\mathcal{C}$ and $\mathfrak{a}$ are as before, and $\mathcal{K}$ is some real function. By rearranging, we write \eqref{necScheme} as
\begin{equation}
	\frac{d}{d\rho}\left[\mathfrak{a}(\rho)^{1/(d-1)}\right]\geq \frac{\mathcal{K}(\rho)^2}{\mathcal{C}(\rho)} \geq 0.
\end{equation}
Thus, if the NEC may be written in this way, we immediately have that $\mathfrak{a}^{1/(d-1)}$ is monotonic. From here, the argument is the same as before; $\mathfrak{a}$ is a monotonic $a$-function if $\mathfrak{a}^{(d-2)/(d-1)} \geq 0$ and $\mathfrak{a} \sim \ell^{d-1}$ in the UV.

As a caveat, while this systematic approach yields a candidate $a$-function which is naturally monotonic in the exterior of the black hole, we are not immediately guaranteed monotonicity in the trans-IR. This is because of ambiguities with signs and factors of $i$. Proving monotonicity in the trans-IR is more easily done in coordinates where the radial direction is real in the interior, e.g. \eqref{coordChange}. Nonetheless, the approach of rewriting the NEC in the form \eqref{necScheme} is still a useful way to \textit{construct} the candidate $a$-function in the first place.

\subsection{Monotonicity in the Exterior} \label{sec:extConstr}

Just as in previous constructions \cite{Freedman:1999gp,Myers:2010tj,Myers:2010xs,Caceres:2022smh}, we consider the radial null vector
\begin{equation}
k^\mu = \frac{e^{-A(\rho)}}{f(\rho)} \delta^\mu_t + \delta^\mu_\rho.
\end{equation}
The corresponding radial NEC is then
\begin{equation}
 (d-1)\left[\frac{f'(\rho)A'(\rho) - f(\rho)A''(\rho)}{f(\rho)}\right] - \delta_1\left[A'(\rho) \mathcal{X}'(\rho) + \mathcal{X}''(\rho) - \frac{f'(\rho)\mathcal{X}'(\rho)}{f(\rho)} + \mathcal{X}'(\rho)^2\right] \ge 0.\label{necAnisoTherm}
\end{equation}
This may be written in the form \eqref{necScheme} by setting
\begin{equation}
\begin{split}
\mathfrak{a}(\rho) &= e^{-\delta_1 \mathcal{X}(\rho)} \left[\frac{(d-1) f(\rho)}{\delta_1 \big(A'(\rho) + \mathcal{X}'(\rho)\big) + \delta_2 A'(\rho)}\right]^{d-1},\\
\mathcal{C}(\rho) &= \frac{1}{(d-1)f(\rho)} \left[\delta_1 \big(A'(\rho) + \mathcal{X}'(\rho)\big) + \delta_2 A'(\rho)\right]^2 e^{\delta_1 \mathcal{X}(\rho)/(d-1)},\\
\mathcal{K}(\rho) &= \sqrt{\frac{\delta_1 \delta_2}{d-1}} \mathcal{X}'(\rho).
\end{split}\label{schematicFunctionsAniso}
\end{equation}
So, unlike in the isotropic cases, the left-hand side of the radial NEC is not just the product of a total derivative and positive function, since $\mathcal{K}$ is not identically $0$. Nonetheless, as discussed in Section \ref{sec:intNEC}, we may still use $\mathfrak{a}$ as a viable candidate $a$-function. For flows of the form \eqref{anisoDomainWall}, we denote this candidate as
\begin{equation}
a_T^{(\delta_1)}(\rho) \sim e^{-\delta_1 \mathcal{X}(\rho)} \left[\frac{(d-1) f(\rho)}{\delta_1 \big(A'(\rho) + \mathcal{X}'(\rho)\big) + \delta_2 A'(\rho)}\right]^{d-1}.\label{monotonicCandidateRho}
\end{equation}
From \eqref{boundaryConds}, this asymptotes to the appropriate UV trace anomaly $\ell^{d-1}$. To ensure that this function is monotonic in $\rho > 0$ of the deformed black hole, we only need to show that
\begin{equation}
e^{-\delta_1 (d-2) \mathcal{X}(\rho)/(d-1)} \left[\frac{(d-1) f(\rho)}{\delta_1 \big(A'(\rho) + \mathcal{X}'(\rho)\big) + \delta_2 A'(\rho)}\right]^{d-2} \geq 0.
\end{equation}
The exponential and $(d-1)f$ factors are always positive in the exterior. Thus we only need to prove that
\begin{equation}
\left[\delta_1 \big(A'(\rho) + \mathcal{X}'(\rho)\big) + \delta_2 A'(\rho)\right]^d \geq 0.
\end{equation}
This is automatically true for even $d$. The case of odd $d$, however, is not as straightforward---we must explicitly show that
\begin{equation}
\delta_1 \big(A'(\rho) + \mathcal{X}'(\rho)\big) + \delta_2 A'(\rho) \geq 0.\label{ineqAniso}
\end{equation}
The proofs discussed in Appendix \ref{app:isoCorrect} may be adjusted to apply here. First, we first note that \eqref{ineqAniso} is already true in the UV by assumption---$A'$ asymptotes to $\ell^{-1} > 0$ while $\mathcal{X}'$ asymptotes to $0$. So then, we assume that there exists a $\rho = \rho_* > 0$ such that the quantity on the left-hand side of \eqref{ineqAniso} is $0$ at $\rho_*$ and positive for $\rho > \rho_*$. By analyticity, we may Taylor-expand,
\begin{equation}
\delta_1 \big(A'(\rho) + \mathcal{X}'(\rho)\big) + \delta_2 A'(\rho) = c_{*} (\rho - \rho_*)^{p_*} + O[(\rho - \rho_*)^{p_*+1}],\label{taylorAniso}
\end{equation}
where $p_* \geq 1$ is the degree of the root while $c_{*} \neq 0$. We now define a small $\varepsilon > 0$ and consider the point $\rho = \rho_* + \varepsilon$. Here, we use the approximation
\begin{equation}
\delta_1 \big(A'(\rho_* + \varepsilon) + \mathcal{X}'(\rho_* + \varepsilon)\big) + \delta_2 A'(\rho_* + \varepsilon) \approx c_* \varepsilon^{p_*},
\end{equation}
to deduce that $c_* > 0$. Next, we differentiate \eqref{taylorAniso} to write
\begin{equation}
\delta_1 \big(A''(\rho) + \mathcal{X}''(\rho)\big) + \delta_2 A''(\rho) = c_{*} p_* (\rho - \rho_*)^{p_*-1} + O[(\rho - \rho_*)^{p_*}].
\end{equation}
By evaluating this at $\rho = \rho_* + \varepsilon$, we have the approximation
\begin{equation}
\delta_1 \big(A''(\rho_*+\varepsilon) + \mathcal{X}''(\rho_*+\varepsilon)\big) + \delta_2 A''(\rho_*+\varepsilon) \approx c_{*} p_* \varepsilon^{p_*-1}.
\end{equation}
We may now approximate the left-hand side of the radial NEC \eqref{necAnisoTherm} at $\rho = \rho_* + \varepsilon$. Specifically, we solve for $A'(\rho_* + \varepsilon)$ and $A''(\rho_* + \varepsilon)$:
\begin{align}
A'(\rho_* + \varepsilon) &\approx \frac{1}{(d-1)}\left[c_* \varepsilon^{p_*} - \delta_1 \mathcal{X}'(\rho_* + \varepsilon)\right],\\
A''(\rho_* + \varepsilon) &\approx \frac{1}{(d-1)}\left[c_* p_* \varepsilon^{p_* - 1} - \delta_1 \mathcal{X}''(\rho_* + \varepsilon)\right].
\end{align}
For notational convenience, let us denote the left-hand side of \eqref{necAnisoTherm} as $\mathcal{N}_\rho$, so the NEC is written as $\mathcal{N}_\rho \ge 0$. The behavior of $\mathcal{N}_\rho$ around $\rho_*$ is
\begin{equation}
\left.\mathcal{N}_\rho\right|_{\rho = \rho_* + \varepsilon} \approx - \frac{\delta_1 \delta_2}{(d-1)}\mathcal{X}'(\rho_* + \varepsilon)^2 - c_* p_* \varepsilon^{p_* - 1} + \varepsilon^{p_*}\left[\frac{c_* f'(\rho_* + \varepsilon)}{f(\rho_* + \varepsilon)} - \frac{c_* \delta_1 \mathcal{X}'(\rho_* + \varepsilon)}{(d-1)}\right].
\end{equation}
Note that we may truncate the $\varepsilon^{p_*}$ term because it is dominated by the $\varepsilon^{p_* - 1}$ term. Mathematically, this means that
\begin{equation}
\left.\mathcal{N}_\rho\right|_{\rho = \rho_* + \varepsilon} \approx - \frac{\delta_1 \delta_2}{(d-1)}\mathcal{X}'(\rho_* + \varepsilon)^2 - c_* p_* \varepsilon^{p_* - 1}.\label{approxFinNEC}
\end{equation}
We have not assumed anything about the root structure of $\mathcal{X}'$ (let alone its square), and so we do not know the order of $\mathcal{X}'(\rho_* + \varepsilon)^2$ in the $\varepsilon$ expansion. However, this does not matter; \eqref{approxFinNEC} is certainly negative and thus represents a violation of the NEC. Thus we have proven that \eqref{ineqAniso} holds and that \eqref{monotonicCandidateRho} is a monotonic function on $\rho > 0$.

\subsubsection*{Anisotropic Vacuum Deformations} 

We may reduce the ansatz \eqref{anisoDomainWall} to one describing anisotropic deformations of the UV vacuum state by setting $f(\rho) = 1$,
\begin{equation}
ds^2 = e^{2A(\rho)}\left[-dt^2 + e^{2\mathcal{X}(\rho)} d\vec{x}^2\right] + d\rho^2.
\end{equation}
In this case, we are left with the monotonic $a$-function,
\begin{equation}
a^{(\delta_1)}(\rho) \sim e^{-\delta_1 \mathcal{X}(\rho)} \left[\frac{(d-1)}{\delta_1\big(A'(\rho) + \mathcal{X}'(\rho)\big) + \delta_2 A'(\rho)}\right]^{d-1}.
\end{equation}
That this is monotonic only requires a single radial NEC. We thus infer that spontaneously broken symmetry away from the UV is not necessarily tied to the monotonicity of holographic RG flow.

Indeed, in a sense we observe that monotonicity becomes \textit{stronger} in the anisotropic case. Specifically, consider the radial NEC (found by setting $f(\rho) = 1$ in \eqref{necAnisoTherm}) in its general form \eqref{necScheme}. Unlike in the isotropic cases, the $\mathcal{K}^2$ term is not uniformly zero; it may be positive. As such, the total-derivative term may be more sharply bounded within the flow.

\subsection{Monotonicity in the Trans-IR} 

We now prove that \eqref{monotonicCandidateRho} is monotonic in the trans-IR, i.e. in the analytic continuation $\rho = i \kappa$ where $\kappa > 0$ \eqref{analyticContCoords}. To do so, we define the coordinate transformation\footnote{We emphasize that our ultimate goal is to bound the derivative of the $a$-function with respect to the $\rho$ coordinate. For technical reasons, we find it more practical to do so in a coordinate system for which the interior radial coordinate runs over $\mathbb{R}$, rather than say $i\mathbb{R}$. The details of this auxiliary coordinate system do not matter so long as we know its Jacobian with $\rho$ coordinate system, which is needed to convert the derivative back to the $\rho$ coordinate.}
\begin{align}
e^{A(\rho)} = \frac{1}{\bar{r}},\ \ f(\rho)^2 = F(\bar{r})e^{-\chi(\bar{r})},\ \ e^{\delta_1 \mathcal{X}(\rho)} = \frac{1}{G(\bar{r})}, \ \ d\rho = -\frac{d\bar{r}}{\bar{r}\sqrt{F(\bar{r})}}.
\end{align}
Implicit in this coordinate transformation is that the UV curvature scale $\ell$ is set to $1$. We take $\chi$ to be some generic real function, $G$ to be some positive function, and $F$ to have a simple root at $\bar{r} = \bar{r}_h$ such that
\begin{equation}
F(\bar{r} < \bar{r}_h) > 0,\ \ F(\bar{r} > \bar{r}_h) < 0.
\end{equation}
Thus, $\bar{r}$ is a real coordinate which covers both the exterior ($\bar{r} < \bar{r}_h$) and interior ($\bar{r} > \bar{r}_h$) of the black hole, just like the radial coordinate defined for the isotropic ansatz via \eqref{coordChange}. However, we reiterate that $\bar{r}$ is not itself the energy scale. The metric \eqref{anisoDomainWall} in these new coordinates takes the form
\begin{equation}
ds^2 = \frac{1}{\bar{r}^2} \left[-F(\bar{r})e^{-\chi(\bar{r})}dt^2 + \frac{d\bar{r}^2}{F(\bar{r})} + \frac{d\vec{x}_1^2}{G(\bar{r})^{2/\delta_1}} + d\vec{x}_2^2\right].\label{coordsReal}
\end{equation}
When $G(\bar{r}) = 1$ (corresponding to $\mathcal{X}(\rho) = 0$), this indeed reduces to the isotropic metric \eqref{isotropic}.

We now write the $a$-function \eqref{monotonicCandidateRho} in these new coordinates. Specifically, by using the chain rule to write
\begin{equation}
A'(\rho) + \frac{\delta_1}{d-1} \mathcal{X}'(\rho) = \sqrt{F(\bar{r})}\left[1 + \frac{\bar{r}}{d-1}\frac{G'(\bar{r})}{G(\bar{r})}\right],
\end{equation}
we deduce that the $a$-function in $\bar{r}$ coordinates is
\begin{equation}
a_T^{(\delta_1)}(\bar{r}) \sim G(\bar{r}) e^{-(d-1)\chi(\bar{r})/2} \left[1 + \frac{\bar{r}}{d-1}\frac{G'(\bar{r})}{G(\bar{r})}\right]^{-(d-1)}.
\end{equation}
Note that the number of warped dimensions $\delta_1$ is implicit in $G$ and does not appear explicitly in this expression.

To show monotonicity everywhere, we ultimately need to prove that
\begin{align}
\text{Exterior $(r < r_h)$:}&\ \ \frac{da_T^{(\delta_1)}}{d\rho} = \frac{d\bar{r}}{d\rho} \frac{da_T^{(\delta_1)}}{d\bar{r}} \geq 0,\label{monExt}\\
\text{Interior $(r > r_h)$:}&\ \ \frac{da_T^{(\delta_1)}}{d\kappa} = \frac{d\rho}{d\kappa}\frac{d\bar{r}}{d\rho}\frac{da_T^{(\delta_1)}}{d\bar{r}} \leq 0.\label{monInt}
\end{align}
The first condition states that the $a$-function decreases from the UV to the IR, while the second condition states that the $a$-function decreases from the IR to the singularity. We have already proven \eqref{monExt} directly in $\rho$ coordinates, but we also check for consistency by proving it in $\bar{r}$ coordinates. We reiterate that we have yet to show \eqref{monInt}.

\subsubsection*{Derivative with Respect to $\bar{r}$}

We first compute the derivative of the $a$-function with respect to $\bar{r}$ and show that the radial NEC demands that it is negative everywhere. We take the radial NEC in \eqref{coordsReal} corresponding to
\begin{equation}
k^\mu = e^{\chi(\bar{r})/2}\delta^\mu_t + F(\bar{r})\delta^\mu_{\bar{r}}.
\end{equation}
The contraction of this against the stress tensor is
\begin{equation}
\mathcal{N}_{\bar{r}} = T_{\mu\nu}k^\mu k^\nu = \frac{F(\bar{r})}{\bar{r}^2}\left(\tensor{T}{^r_r} - \tensor{T}{^t_t}\right).
\end{equation}
By computing the stress tensor explicitly, solving for $G''$ in terms of $\mathcal{N}_{\bar{r}}$, and plugging its solution into the derivative of the $a$-function with respect to $\bar{r}$, we find that
\begin{equation}
\frac{da_T^{(\delta_1)}}{d\bar{r}} \sim -\frac{e^{-(d-1)\chi(\bar{r})/2}\bar{r} G(\bar{r})}{F(\bar{r})^2} \left[1 + \frac{\bar{r}}{d-1}\frac{G'(\bar{r})}{G(\bar{r})}\right]^{-d} \left[\mathcal{N}_{\bar{r}} + \frac{\delta_2}{(d-1)\delta_1} \left(\frac{G'(\bar{r})}{G(\bar{r})}\right)^2 F(\bar{r})^2\right].
\end{equation}
This expression may also be reproduced by applying a coordinate transformation to the left-hand side of \eqref{necScheme}, which we denote as $\mathcal{N}_\rho$. In doing so, we must be careful to note that $\mathcal{N}_\rho$ and $\mathcal{N}_{\bar{r}}$ are computed by null vectors with different overall scales and are thus related as
\begin{equation}
\mathcal{N}_\rho = \frac{\bar{r}^2}{F(\bar{r})} \mathcal{N}_{\bar{r}}.
\end{equation}
As a further sanity check, we may set $G(\bar{r}) = 1$ in order to reproduce both the $a$-function and the derivative in the isotropic case \cite{Caceres:2022smh}.

This derivative is proportional to the quantity $\mathcal{N}_{\bar{r}}$ bounded by the NEC plus a square, and this sum must be nonnegative. However, unlike in the isotropic case, there is an ambiguous factor in front. Specifically, we know that
\begin{equation}
\left[1 + \frac{\bar{r}}{d-1}\frac{G'(\bar{r})}{G(\bar{r})}\right]^{d} \frac{da_T^{(\delta_1)}}{d\bar{r}} \sim -\frac{e^{-(d-1)\chi(\bar{r})/2}\bar{r} G(\bar{r})}{F(\bar{r})^2} \left[\mathcal{N}_{\bar{r}} + \frac{\delta_2}{(d-1)\delta_1} \left(\frac{G'(\bar{r})}{G(\bar{r})}\right)^2 F(\bar{r})^2\right] \leq 0,
\end{equation}
by the radial NEC and the positivity of $\bar{r}$, $G$, $F^2$, and the exponential of $\chi$, but we still need to prove that
\begin{equation}
1 + \frac{\bar{r}}{d-1} \frac{G'(\bar{r})}{G(\bar{r})} \geq 0,
\end{equation}
everywhere in the geometry. This follows from the asymptotics of the expression and the radial NEC, just as it does for the analogous expression in $\rho$ coordinates \eqref{ineqAniso}, except this time the proof applies both to the exterior and interior regions of the black hole.

Indeed, following the same procedure as before, we start by noting that
\begin{equation}
\lim_{\bar{r} \to 0} \left[1 + \frac{\bar{r}}{d-1} \frac{G'(\bar{r})}{G(\bar{r})}\right] = 1 > 0.
\end{equation}
Furthermore, we assume that there is an $\bar{r} = \bar{r}_*$ for which the expression vanishes and \textit{below} which the expression is positive. By defining a small $\varepsilon > 0$, we may approximate to leading-order in $\varepsilon$:
\begin{align}
&1 + \frac{\bar{r}_* - \varepsilon}{d-1} \left[\frac{G'(\bar{r}_* - \varepsilon)}{G(\bar{r}_* - \varepsilon)}\right] \approx c_* (-\varepsilon)^{p_*},\label{subepsilon}\\
&\frac{1}{d-1}\left[\frac{G'(\bar{r}_* - \varepsilon)}{G(\bar{r}_* - \varepsilon)} + \frac{(\bar{r}_* - \varepsilon) G''(\bar{r}_* - \varepsilon)}{G(\bar{r}_* - \varepsilon)} - \frac{(\bar{r}_* - \varepsilon)G'(\bar{r}_* - \varepsilon)^2}{G(\bar{r}_* - \varepsilon)^2}\right] \approx c_* p_* (-\varepsilon)^{p_* - 1},
\end{align}
where $c_* \neq 0$ and $p_* \geq 1$. By solving for $G'(\bar{r}_* - \varepsilon)$ and $G''(\bar{r}_* - \varepsilon)$ and subsequently plugging into the radial NEC, we get (after dropping terms which are certainly subleading)
\begin{equation}
\left.\mathcal{N}_{\bar{r}}\right|_{\bar{r} = \bar{r}_* - \varepsilon} \approx -\frac{(d-1)F(\bar{r}_*)^2}{\bar{r}_*^2}\left[\frac{\delta_2}{\delta_1} - c_* p_* \bar{r}_* (-\varepsilon)^{p_* - 1}\right].\label{necscalarReal}
\end{equation}
If $p_* > 1$, then we find that $\mathcal{N}_{\bar{r}}$ is negative and we have a contradiction. If $p_* = 1$, then we must use the fact that the left-hand side of \eqref{subepsilon} is positive by assumption to state that $c_* < 0$ when $p_* = 1$, so we still conclude that $\mathcal{N}_{\bar{r}}$ is negative even in this case.

There is one edge case. Note that the expression \eqref{necscalarReal} is actually $0$ if $\bar{r}_* = \bar{r}_h$, and thus our argument breaks down here. However, we may simply rescale the null vector and, thus, $\mathcal{N}_{\bar{r}}$ so there is no overall factor of $F$ to contend with in the first place.

So, to summarize, using analogous arguments to those of the exterior of the domain-wall ansatz, we have proven that
\begin{equation}
1 + \frac{\bar{r}}{d-1} \frac{G'(\bar{r})}{G(\bar{r})} \geq 0,
\end{equation}
everywhere in the black hole. It follows that
\begin{equation}
\frac{da_T^{(\delta_1)}}{d\bar{r}} \leq 0.\label{ineqAr}
\end{equation}

\subsubsection*{Chain Rule}

We are now ready to show \eqref{monExt}--\eqref{monInt}. We simply use the chain rule in conjunction with \eqref{ineqAr} to do so. For the exterior $\bar{r} < \bar{r}_h$, we first write
\begin{equation}
\frac{d\bar{r}}{d\rho} = -\bar{r}\sqrt{F(\bar{r})} \leq 0 \implies \left.\frac{da_T^{(\delta_1)}}{d\rho}\right|_{\bar{r} < \bar{r}_h} \geq 0,
\end{equation}
in agreement with the general argument employed in \eqref{sec:extConstr}. As for the interior $\bar{r} > \bar{r}_h$, however, we must recall that \eqref{analyticContCoords} to write
\begin{equation}
\frac{d\rho}{d\kappa} = i \implies \frac{d\rho}{d\kappa}\frac{d\bar{r}}{d\rho} = -i\bar{r}\sqrt{F(\bar{r})} = \bar{r}\sqrt{|F(\bar{r})|} \geq 0.
\end{equation}
This is because $F$ is \textit{negative} in the interior, and so it produces an additional factor of $i$. Thus, we have that
\begin{equation}
\left.\frac{da_T^{(\delta_1)}}{d\kappa}\right|_{r > r_h} \leq 0.
\end{equation}
We conclude that $a_T^{(\delta_1)}$ is monotonic along the full flow, including the trans-IR.

As an aside, observe that employing the chain rule at the horizon allows us to state that the derivative of the $a$-function vanishes identically at the horizon $\bar{r} = \bar{r}_h$. This is consistent with the horizon being the IR fixed point.

\section{Holographic $p$-Wave Superfluids} \label{sec:pwave}

As a simple test case, we consider the anisotropic $p$-wave superfluid solution of \cite{Ammon:2009xh}. This is a concrete example of a holographic flow geometry exhibiting aniostropy, and so we may study an explicit $a$-function. In keeping with their conventions, we set $d = 4$ in this section. We thus consider a $(4+1)$-dimensional Einstein--Yang--Mills theory with $SU(2)$ gauge symmetry whose bulk action is (setting $\ell = 1$)
\begin{equation}
I_{\text{EYM}} = \int d^5 x\sqrt{-g} \left[\frac{1}{2\ell_{\text{P}}^3} \left(R + 12\right) - \frac{1}{4\hat{g}^2} F^{a}_{\mu\nu} F^{a\mu\nu}\right].\label{actEYM}
\end{equation}
$\hat{g}$ is the Yang--Mills coupling while $a = 1,2,3$ runs over the generators of the (three-dimensional representation of the) $SU(2)$ gauge group. These generators are denoted as $\tau^a \equiv \dfrac{\sigma^a}{2i}$, where $\{\sigma^1,\sigma^2,\sigma^3\}$ are the Pauli matrices. $F_{\mu\nu}^a$ is the $SU(2)$ field strength related to the gauge field $\mathcal{A}^a_{\mu}$ by
\begin{equation}
F_{\mu\nu}^a = \partial_\mu \mathcal{A}^a_\nu - \partial_\nu \mathcal{A}^a_\mu + \epsilon_{abc}\mathcal{A}^b_\mu \mathcal{A}^c_\nu,
\end{equation}
and with the Levi-Civita normalized as $\epsilon_{123} = +1$. The full equations of motion are then \cite{Ammon:2009xh}
\begin{align}
R_{\mu\nu} + 4g_{\mu\nu} &= \ell_{\text{P}}^3 \left(\tilde{T}_{\mu\nu} - \frac{1}{3}\tensor{\tilde{T}}{^\sigma_\sigma} g_{\mu\nu}\right),\label{eomEYM1}\\
\nabla_\mu F^{a\mu\nu} &= -\epsilon_{abc} \mathcal{A}^b_\mu F^{c\mu\nu},\label{eomEYM2}
\end{align}
where $\tilde{T}_{\mu\nu}$ is
\begin{equation}
\tilde{T}_{\mu\nu} = \frac{1}{\hat{g}^2} \Tr_{SU(2)}\left(F_{\sigma\mu} \tensor{F}{^\sigma_\nu} - \frac{1}{4} g_{\mu\nu} F_{\sigma\rho}F^{\sigma\rho}\right).
\end{equation}
The class of solutions with which we are concerned is found by using the ansatz,
\begin{align}
ds^2 &= -N(r)\sigma(r)^2 dt^2 + \frac{1}{N(r)}dr^2 + r^2 h(r)^{-4} dx^2 + r^2 h(r)^2 (dy^2 + dz^2),\label{metricpwave}\\
\mathcal{A} &= \phi(r) \tau^3 dt + w(r)\tau^1 dx,\label{oneform}
\end{align}
where $r = \infty$ is the conformal boundary, $r = r_h$ denotes the horizon (at which $N$ has a simple pole), and $r = 0$ is the singularity. Observe that \eqref{metricpwave} is an anisotropic solution. Note that this $r$ is not the same as $\bar{r}$.

Defining a function $m(r)$ by $N(r) = -\dfrac{2m(r)}{r^2} + r^2$, \cite{Ammon:2009xh} rewrites the independent components of the equations of motion in terms of the functions $\{m,\sigma,h,\phi,w\}$ (using primes to denote derivatives and defining a constant $\alpha \equiv \dfrac{\ell_{\text{P}}^{3/2}}{\hat{g}}$):
\begin{align}
m' &= \frac{\alpha^2 r h^4 w^2 \phi^2}{6N\sigma^2} + \frac{\alpha^2 r^3 \phi'^2}{6\sigma^2} + N\left(\frac{r^3 h'^2}{h^2} + \frac{\alpha^2 r h^4 w'^2}{6}\right),\label{eompwave1}\\
\sigma' &= \frac{\alpha^2 h^4 w^2 \phi^2}{3rN^2 \sigma} + \sigma\left(\frac{2rh'^2}{h^2} + \frac{\alpha^2 h^4 w'^2}{3r}\right),\label{eompwave2}\\
h'' &= -\frac{\alpha^2 h^5 w^2 \phi^2}{3r^2 N^2 \sigma^2} + \frac{\alpha^2 h^5 w'^2}{3r^2} - h'\left(\frac{3}{r} - \frac{h'}{h} + \frac{N'}{N} + \frac{\sigma'}{\sigma}\right),\label{eompwave3}\\
\phi'' &= \frac{h^4 w^2 \phi}{r^2 N} - \phi'\left(\frac{3}{r}+ \frac{\sigma'}{\sigma}\right),\label{eompwave4}\\
w'' &= -\frac{w\phi^2}{N^2\sigma^2} - w'\left(\frac{1}{r} + \frac{4h'}{h} + \frac{N'}{N} + \frac{\sigma'}{\sigma}\right).\label{eompwave5}
\end{align}
We note that the above ansatz indeed satisfies the radial NEC. This can be shown by employing the equations of motion. The details are discussed in Appendix \ref{app:necpwave}.

\subsection{Writing the Anisotropic $a$-Function}

We write the $a$-function corresponding to the metric \eqref{metricpwave}. The first step is to transform \eqref{metricpwave} into a domain-wall slicing. We use the coordinate transformation $r \to \rho$:
\begin{equation}
\frac{dr}{\sqrt{N(r)}} = d\rho,\ \ N(r)\sigma(r)^2 = e^{2A(\rho)}f(\rho)^2,\ \ r^2 h(r)^{-4} = e^{2[A(\rho) + \mathcal{X}(\rho)]},\ \ r^2 h(r)^2 = e^{2A(\rho)}.\label{coordsTransReal}
\end{equation}
This yields the anisotropic domain-wall metric \eqref{anisoDomainWall} with $\delta_1 = 1$ and $\delta_2 = 2$,
\begin{equation}
ds^2 = e^{2A(\rho)} \left[-f(\rho)^2 dt^2 + e^{2\mathcal{X}(\rho)} dx^2 + dy^2 + dz^2\right] + d\rho^2.\label{aniso21}
\end{equation}
In these coordinates, we know that the $a$-function is
\begin{equation}
a_T^{(1)}(\rho) \sim e^{-\mathcal{X}(\rho)} \left[\frac{f(\rho)}{A'(\rho) + \frac{1}{3}\mathcal{X}'(\rho)}\right]^3.\label{mon12}
\end{equation}
So now we convert back to $r$ coordinates. We find that
\begin{equation}
A'(\rho) + \frac{1}{3}\mathcal{X}'(\rho) = \frac{\sqrt{N(r)}}{r},\ \ f(\rho) = \frac{\sqrt{N(r)}\sigma(r)}{r h(r)},\ \ e^{-\mathcal{X}(\rho)} = h(r)^3.
\end{equation}
Combining all of this, we write
\begin{equation}
a_T^{(1)}(r) \sim \sigma(r)^3.\label{afunc}
\end{equation}
Furthermore, we may write the derivatives of the $a$-function with respect to energy in both the exterior and interior. Defining $\Lambda$ to represent $\rho$ in the exterior and $\kappa$ in the interior, we write
\begin{align}
\frac{da_T^{(1)}}{d\Lambda} \sim \begin{cases}
3\sigma(r)^2 \sigma'(r) \sqrt{N(r)},\ \ &\text{if}\ r > r_h,\\
-3\sigma(r)^2 \sigma'(r) \sqrt{|N(r)|},\ \ &\text{if}\ r < r_h.
\end{cases}\label{afuncDeriv}
\end{align}

\subsection{The $a$-Function in a $p$-wave Superfluid}

Our purpose for employing the $p$-wave superfluid is to concretely observe the behavior of our $a$-function in a single known solution. To this end, we only seek to reconstruct the solution shown in Figure 1 of \cite{Ammon:2009xh}. See Appendix \ref{app:numerics} for technical details regarding how the solutions are generally constructed. First, we note that the physical parameters used by \cite{Ammon:2009xh} are
\begin{equation}
r_h = 1,\ \ \alpha = 0.316.
\end{equation}
\begin{figure}
\centering
\subfloat[Exterior $a$-function]{
\includegraphics[scale=0.55]{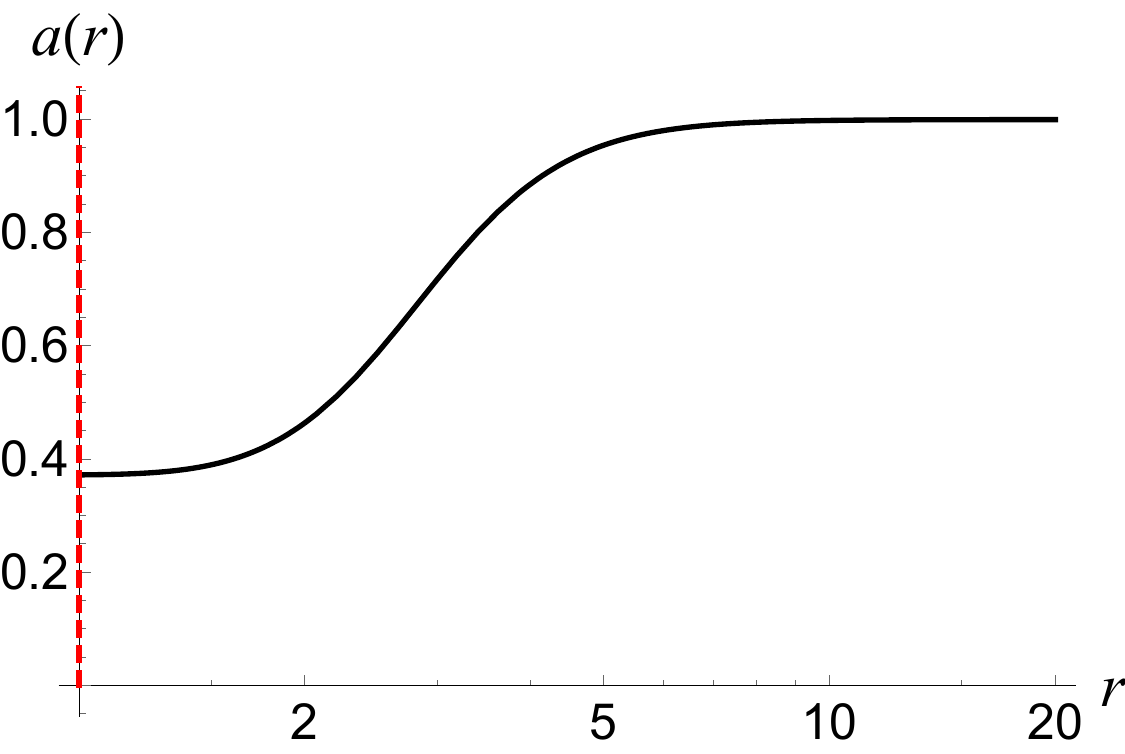}
}\qquad
\subfloat[Interior $a$-function]{
\includegraphics[scale=0.55]{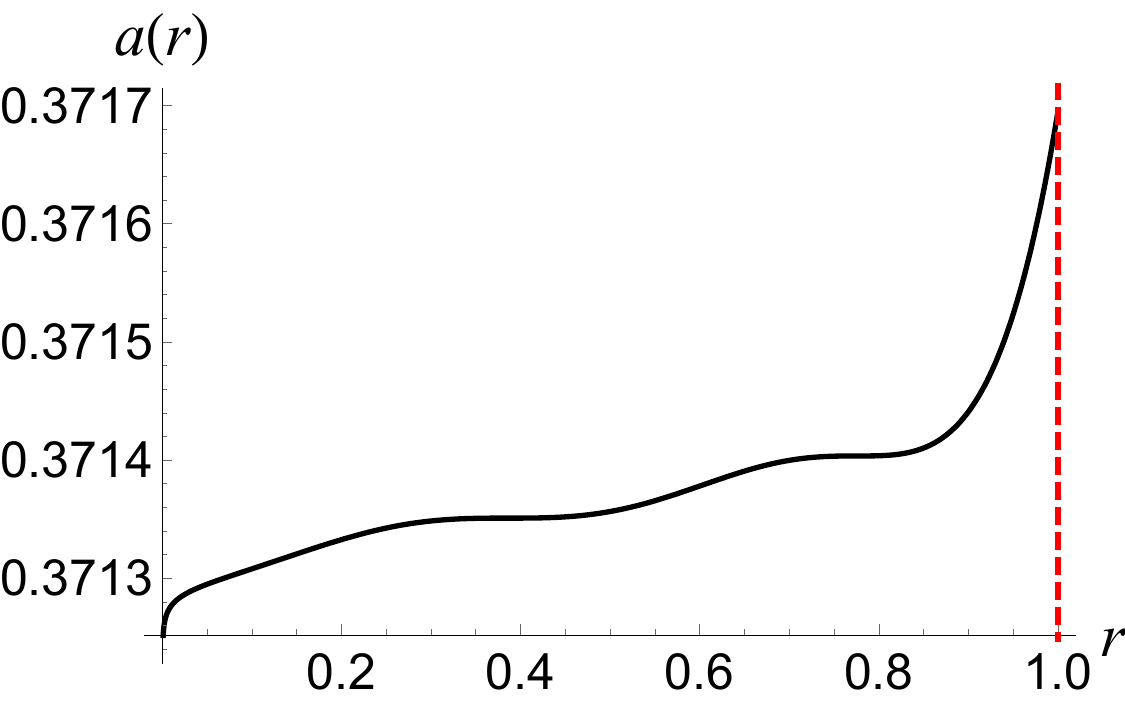}
}
\caption{Plots of the $a$-function \eqref{afunc} in the holographic $p$-wave superfluid state of \cite{Ammon:2009xh}, in terms of $r$ (in units $r_h = 1$). In the exterior (a), the function is relatively constant until $r < 10$, at which point it sharply declines prior to reaching the horizon (in red). Within the interior (b), the function then exhibits fluctuations as it continues decreasing.}
\label{figs:aPlot}
\end{figure}
\begin{figure}
\centering
\subfloat[Exterior $a$-function derivative $\left(\dfrac{da_T^{(1)}}{d\rho}\right)$]{
\includegraphics[scale=0.55]{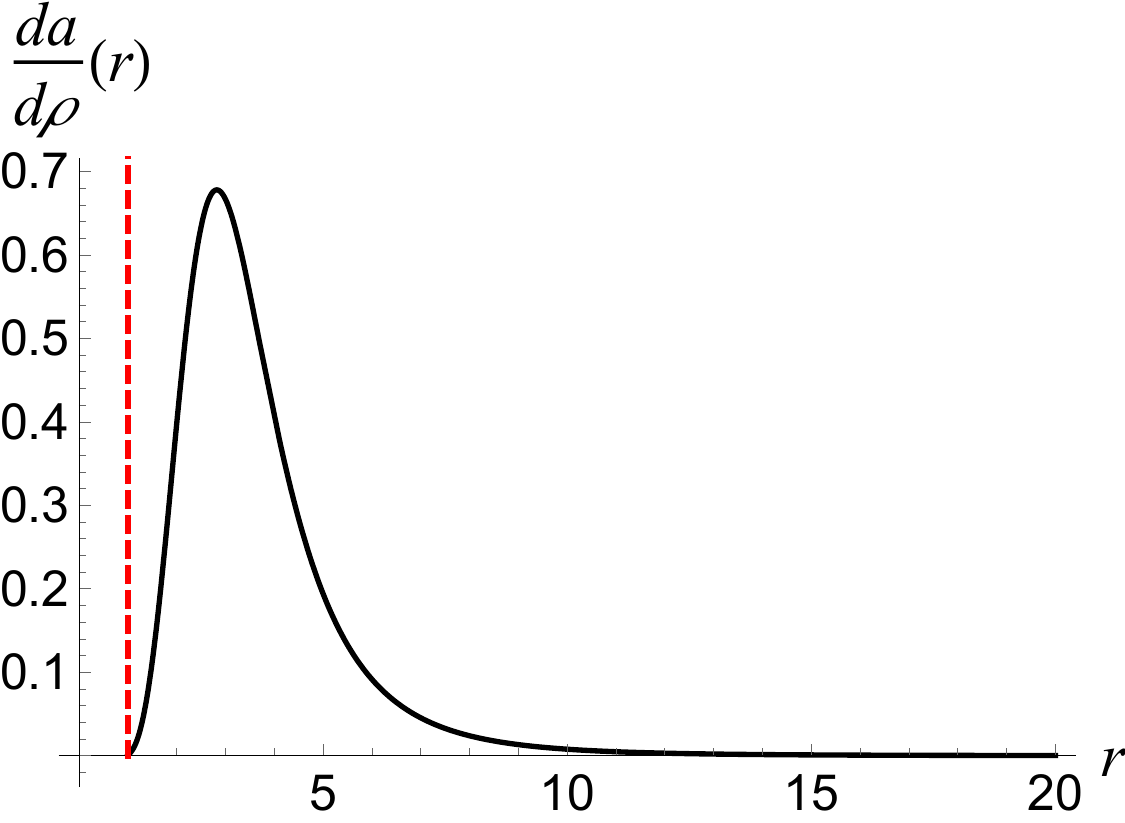}
}\qquad
\subfloat[Interior $a$-function derivative $\left(\dfrac{da_T^{(1)}}{d\kappa}\right)$]{
\includegraphics[scale=0.55]{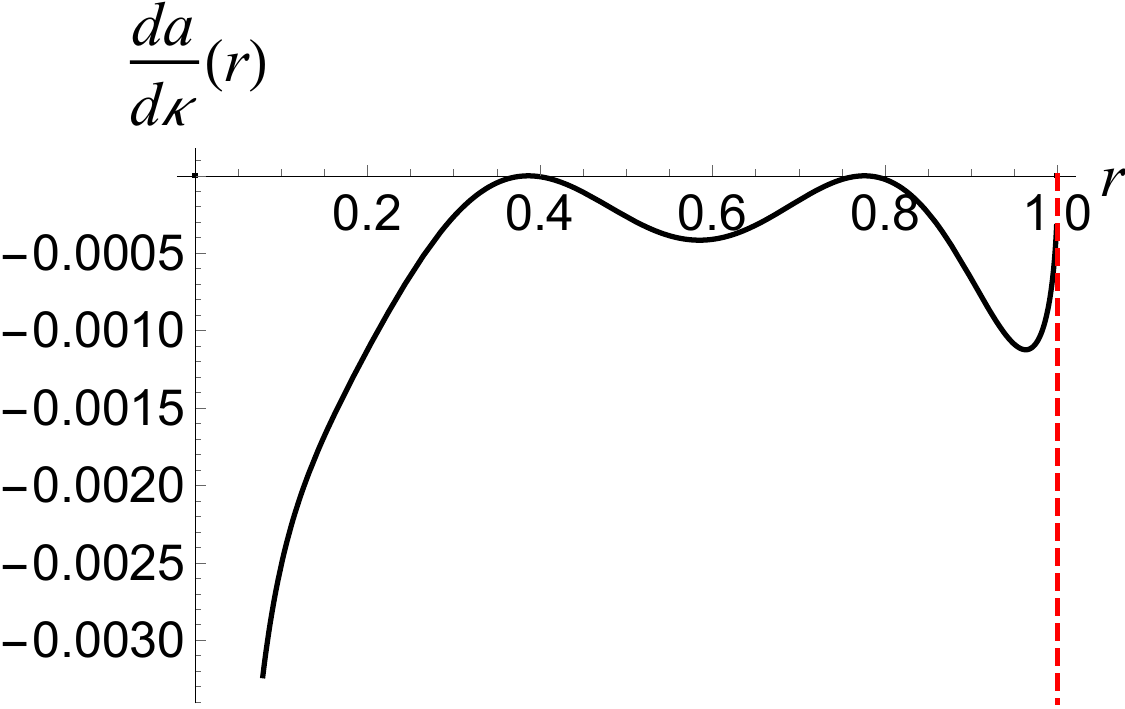}
}
\caption{Plots of the derivative of the $a$-function with respect to the energy scale \eqref{afuncDeriv} in the holographic $p$-wave superfluid state of \cite{Ammon:2009xh}. In the exterior ($\Lambda = \rho$), the derivative is positive, so the $a$-function indeed monotonicially decreases with energy. Furthermore, the derivative goes to $0$ both at $r \to \infty$ and $r = r_h$, in agreement with these being, respectively, the UV and IR fixed points. In the interior ($\Lambda = \kappa$), note that the derivative does not obviously go to $0$ at $r = r_h$ due to numerical instability, but it is negative as expected. Additionally, it undergoes nontrivial oscillations before apparently diverging, unlike in the free Kasner flows of \cite{Caceres:2022smh}. It is possible that the local maxima are fixed points of the trans-IR flow.}
\label{figs:aDerivPlot}
\end{figure}

\noindent When shooting from the horizon, we perform series expansions of the functions $\{m,\sigma,h,\phi,w\}$ about $r = r_h$ \eqref{series1}--\eqref{series5}. However, both general physical reasoning and the equations of motion imply that there are ultimately four free coefficients which we call the ``shooting parameters." These are $\{\sigma_0,h_0,w_0,\phi_1\}$. From Figure 1 in \cite{Ammon:2009xh}, we read off the first three shooting parameters as
\begin{equation}
w_0 = 6.8,\ \ h_0 = 1.061,\ \ \sigma_0 = 0.719.\label{paramsShooting1}
\end{equation}
This leaves $\phi_1$. However, since we are working with flows which are asymptotically isotropic and AdS in the UV, we know that
\begin{align}
\lim_{r\to\infty} h(r) = \lim_{r\to\infty} = \sigma(r) = 1.
\end{align}
This only happens given the above parameters for a specific value of $\phi_1$. We find this to be
\begin{equation}
\phi_1 \approx 0.131.
\end{equation}
Thus, we have enough information to construct a black hole solution to the Einstein--Yang--Mills theory \eqref{actEYM} which may be seen as an anisotropic RG flow.

With these parameters, we perform the shooting both in the exterior as validation (see Appendix \ref{app:ext}) and the interior which is not done by \cite{Ammon:2009xh} (see Appendix \ref{app:int}) using the initial conditions \eqref{initConds}. With these functions in hand, we are ready to plot both the $a$-function \eqref{afunc} and its derivative \eqref{afuncDeriv}. These are shown in Figure \ref{figs:aPlot} and Figure \ref{figs:aDerivPlot}, respectively.

These plots satisfy a variety of sanity checks. For instance, we have been working with a normalization of the $a$-function whereby it is $1$ in the UV, and we see that reflected in our plot. Furthermore, the plots are consistent with the analytic statements that the derivative with respect to the energy scale are $0$ at the horizon and as $r \to \infty$. In other words, the plots are consistent with $r = r_h$ and $r = \infty$ being, respectively, the IR and UV fixed points.

Nonetheless, this $a$-function is less trivial than those seen in free Kasner flows \cite{Caceres:2022smh}, in which the bulk matter sector only consists of a free massive scalar field. Indeed, we see fluctuations in the $a$-function which leave an imprint on the derivative. This and the other fluctuations in the interior functions (Appendix \ref{app:int}) are reminiscent of the Josephson oscillations in the holographic superconductor \cite{Hartnoll:2020fhc,An:2022lvo}. Furthermore, our numerics suggest that such oscillations in the derivative can allow it to reach $0$. This serves as preliminary evidence that oscillations in the interior may drive trans-IR flows into additional nontrivial fixed points. However, we emphasize this an analytic demonstration of this phenomenon is still necessary. That being said, we leave further exploration of interior oscillatory phenomena and their imprint on the holographic $a$-function to future work.
\vfill

\section{Discussion} \label{sec:disc}

To summarize, although we work in flows which spontaneously break rotational symmetry and are thus invariant, we still find that a single NEC corresponding to positivity of energy along the flow direction is sufficient to realize a monotonic $a$-function. This includes holographic trans-IR flows, providing evidence that the $a$-function is a robust means of probing the physics of black hole interiors.

Indeed, this work includes the first application of our holographic $a$-function to a rather nontrivial $p$-wave superfluid state. Unlike in free Kasner flows \cite{Frenkel:2020ysx,Caceres:2022smh}, we may have expected that the nontrivial dynamics in the interior of this black hole would be reflected in the $a$-function. We have shown this expectation to be true.

Within the context of modern AdS/CFT, there are two general lines of research to which our work connects: the general interpretation of holographic RG flow in the context of black hole states and the quantum-information approach to the black hole interior. The statements we make have implications for both.

\subsection{Mysteries of the Trans-IR Flow Interpretation}

In the usual story of holographic RG flow, the classical Hamilton-Jacobi dynamics of gravity precisely correspond to renormalization group (RG) equations dictating the coarse graining of some lower-dimensional quantum state. For example, Hamilton-Jacobi recovers the Callan-Symanzik equations when considering 5-dimensional supergravity and a dual 4-dimensional large-$N$ gauge theory \cite{deBoer:1999tgo}. Imposing the radial NEC essentially means that classical energy density is positive along this flow, so intuitively we may interpret the radial NEC as a unitarity condition for the states on which the RG equations act, in line with \cite{Hofman:2008ar,Kulaxizi:2010jt}. Since the radial NEC implies the existence of a monotonic $a$-function, it makes sense to interpret the $a$-function as a count of degrees of freedom.

However, whether this is the right way to think about holographic trans-IR flows is not as clear. While we have the notion of a monotonic $a$-function, we do not have a clear idea of what the states or even the RG equations are. In other words, while we may say that the trans-IR flow describes some coarse graining if we take seriously the $a$-function as a count of degrees of freedom, it is not clear \textit{what} is being coarse-grained.

There has been some recent progress; \cite{Hartnoll:2022snh} argues that the appropriate analog to the Hamilton-Jacobi is the Wheeler-DeWitt equation in the interior. As such, we may ask if analytic continuations of $a$-functions to the black hole interior describe a coarse graining of Wheeler-DeWitt states. It would also be interesting to formulate path integral complexity \cite{Caputa:2021pad} using these states and see if that may also capture the degrees of freedom along the trans-IR flow.

One complication is seen in the $p$-wave superfluid solution above; it is not clear if the $a$-function goes to $0$. This runs counter to the discussion in \cite{Caceres:2022smh} (supported by numerically constructed free Kasner flows) in which we had argued that the $a$-function should reach $0$ for Belinskii-Khalatnikov-Lifshitz (BKL) singularities \cite{Lifshitz:1963ps,Belinskii:1970ew,Belinskii:1982pk}. There are two possibilities: either the numerical $p$-wave superfluid solution is not reliable close to the singularity, or there is more subtlety concerning when the $a$-function does or does not go to $0$ at the singularity. A more fundamental understanding of what exactly the $a$-function is counting may provide additional insight into this problem.

\subsection{Quantum Information and Trans-IR Flows}

The quantum information of the UV state is commonly viewed as a framework with which to think about the interior of the black hole \cite{Balasubramanian:1999zv,Fidkowski:2003nf,Ryu:2006bv,Hubeny:2007xt,Hartman:2013qma,Susskind:2014rva,Stanford:2014jda,Susskind:2014moa,Brown:2015bva,Brown:2015lvg,Belin:2021bga}. Since we are proposing that the language of trans-IR flows is also such a framework, it behooves us to connect our picture to the various quantum-information-theoretic results.

In \cite{Caceres:2022smh}, we do so by asking how far the holographic geometric duals of various information-theoretic measures probe into the interior. This provides an alternate way to phrase the statement that classical entanglement \cite{Ryu:2006bv,Hubeny:2007xt,Hartman:2013qma} and complexity from volume \cite{Susskind:2014rva,Stanford:2014jda,Susskind:2014moa} are generally ``not enough" to probe the full interior \cite{Susskind:2014moa}. Specifically, whenever we have extremal surface ``barriers" (which are believed to be generic features of AdS gravity \cite{Engelhardt:2013tra}), both prescriptions only cover a finite part of the trans-IR flow and never probe close to the singularity. Meanwhile, geodesic approximation of 2-point functions \cite{Balasubramanian:1999zv,Fidkowski:2003nf} and complexity from the action of the Wheeler-DeWitt patch \cite{Brown:2015bva,Brown:2015lvg} \textit{are} sufficient to probe the full interior, because the associated geometric objects are capable of reaching the singularity.

Our arguments in \cite{Caceres:2022smh} were exemplified by relatively simple free Kasner flows. However, we may ask whether a similar story may be told for more complicated black holes, such as those with charge \cite{Carmi:2017jqz} or anisotropy. There has been some work on this; \cite{Auzzi:2022bfd} considers such charged black holes with backreaction, admitting an RG flow interpretation. It would be interesting to more concretely interpret their results in the language of holographic RG flow, radial NECs, and $a$-functions.

More tangibly to the point of making contact with quantum information, recall that entanglement entropy is associated with the holographic $a$-function in Lorentzian flows \cite{Myers:2010tj,Myers:2010xs}. However, it is known that entanglement entropy is not guaranteed to be monotonic along Lorentz-violating flows \cite{Swingle:2013zla} unless additional constraints are applied \cite{Cremonini:2013ipa}. It is thus reasonable to ask whether entanglement is truly the quantum-information quantity related to the $a$-function. As a first step, one may ask if the holographic flows considered here always satisfy the additional constraints of \cite{Cremonini:2013ipa} which imply monotonicity of entanglement entropy.

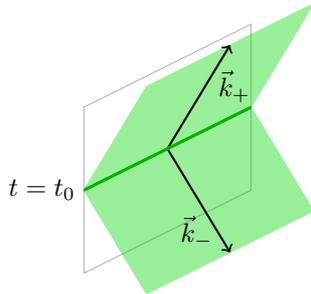
\begin{figure}
\centering
\begin{tikzpicture}[scale=1.1]

\draw[-,black!30] (0,0) to (0,2) to (2,3) to (2,1) to (0,0);

\draw[-,black!15!green,fill=black!15!green,opacity=0.4] (0,1) to (0.75,2.25) to (2.75,3.25) to (2,2);
\draw[-,black!15!green,fill=black!15!green,opacity=0.4] (0,1) to (0.75,-0.25) to (2.75,0.75) to (2,2);

\draw[->,thick] (1,1.5) to (1.75,2.75);
\draw[->,thick] (1,1.5) to (1.75,0.25);

\node at (1.8,2.25) {$\vec{k}_+$};
\node at (1.35,0.5) {$\vec{k}_-$};

\draw[-,black!30!green,very thick] (0,1) to (2,2);
\node at (-0.5,1) {$t = t_0$};

\end{tikzpicture}
\caption{The Wheeler-DeWitt patch corresponding to some constant boundary-time slice $t = t_0$. The radial NEC corresponding to the null vectors $\vec{k}_{\pm}$ produces a monotonic $a$-function.}
\label{figs:wdwpatchNEC}
\end{figure}

Nonetheless, whether or not Lorentz symmetry is spontaneously broken, we may take the radial NEC as our starting point. This condition can be related to the Wheeler-DeWitt patch (Figure \ref{figs:wdwpatchNEC}). To see how, observe that the patch whose action computes holographic complexity \cite{Brown:2015bva,Brown:2015lvg} at some boundary time $t = t_0$ is defined by null sheets anchored at this boundary-time slice. The tangent vectors of the null sheets are precisely the sort of null vectors needed in the radial NEC. Thus, one way to state our conjecture regarding monotonicity (Section \ref{sec:radNEC}) is as follows:
\begin{quote}
\textit{If we have energy positivity along the null sheets bounding the Wheeler-DeWitt patches which compute complexity as a function of boundary time, the we have a monotonic function along the full flow.}
\end{quote}
Since Wheeler-DeWitt patches cross black hole horizons, this statement applies to trans-IR flows. Furthermore, we may claim that it is really holographic complexity from action that properly captures the holographic $a$-function. We leave more precise explorations of this idea to future work.

\subsection{Other Future Directions}

There are other concrete avenues for future work. We list some of these below.

\paragraph{Phase transitions} It would be interesting to conduct a more thorough analysis of holographic $p$-wave superfluid states \cite{Ammon:2009xh,Cai:2021obq} so as to see how the $a$-function may be sensitive to the phase structure of the underlying theory. One may also consider the $a$-function in other setups exhibiting phase transitions, such as the holographic superconductor \cite{Hartnoll:2020fhc,An:2022lvo} or round black holes \cite{Caceres:2022pts}.

\paragraph{Increased symmetry breaking} The anisotropic ansatz studied in this paper is a rather minimal model of symmetry breaking which still maintains two rotationally invariant subspaces \eqref{symmBreak}. We may ask whether the story of monotonicity being guaranteed by the radial NEC still goes through for ans\"atze that further break Lorentz invariance. We expect this to be the case, but it would be satisfying to systematically demonstrate as much in a more general class of flows.

\paragraph{Breaking translation symmetry} We may also study holographic configurations with other types of spontaneous symmetry breaking. For example, we may consider flows which break translation invariance and thus describe phonons. Simple flows of this kind include those for which the dynamics are governed by ODEs \cite{Baggioli:2014roa,Baggioli:2021xuv,Baggioli:2022aft,Baggioli:2022pyb,Liu:2022rsy}. We may also take the striped superconductor \cite{Donos:2011bh,Donos:2013gda}, but this is more complicated because the dynamics are governed by PDEs and would thus require more powerful numerical techniques. Nonetheless, our approach to deriving the $a$-function should go through for ans\"atze supporting broken translational invariance in the bulk.

\subsection{Conclusions}

In conclusion, even with reduced symmetry, the monotonicity of our holographic $a$-function is rooted in the assumption of just the radial null energy condition---even in the interior of black holes. While we interpret such a condition as the imposition of energy positivity along the holographic RG flow direction, there are still mysteries concerning what exactly is flowing in the trans-IR. Nonetheless, even if the trans-IR flow does not have an interpretation as a coarse-graining flow between states, our $a$-function is a well-defined and general gravitational construct which has access to the singularity and thus warrants further study.

\acknowledgments 

We are grateful to Ted Jacobson for asking the question initiating this work. We also thank Oscar Sumari Barron for collaboration during the early stages of this work and Aaron Zimmerman for useful discussion. We are also grateful to Matteo Baggioli, Dimitrios Giataganas, Arnab Kundu, and Juan Pedraza for comments on the draft. EC thanks the Galileo Galilei Institute for Theoretical Physics for hospitality and the participants of the workshop ``Reconstructing the Gravitational Hologram with Quantum Information" for discussions. EC and SS are supported by National Science Foundation (NSF) Grant No. PHY-2112725. SS is also supported by National Science Foundation (NSF) Grant No. PHY-1914679.

\vfill
\pagebreak

\begin{appendices}

\section{Corrected Proof for Isotropic Deformations}\label{app:isoCorrect} 

In this section, we provide more complete analytic proofs of the positivity of the metric functions required for monotonicity of $a$-functions in odd $d$. We focus on the isotropic deformations, but the general technique used herein may also be applied to the anistropic cases (see Section \ref{sec:extConstr}). In the language of Section \ref{sec:intNEC}, these are proofs that $\mathfrak{a}(\rho)^{(d-2)/(d-1)} \geq 0$ when such a term is not already a square.

Regarding our assumptions, we always take the metric functions to be analytic and smooth on spacetime and to have behavior as $\rho \to \infty$ such that the metric is asymptotically AdS. We also work in the domain $\rho > 0$ and assume that the constant-$\rho$ slices are Lorentzian, so these proofs are not enough to prove monotonicity in trans-IR regimes. Lastly, we only require the radial NEC.

\subsection{Vacuum Deformations}\label{app:isoVac}

Given the metric \eqref{vacmet},
\begin{equation}
ds^2 = e^{2A(\rho)}\left(-dt^2 + d\vec{x}^2\right) + d\rho^2,
\end{equation}
we wish to show that $A'(\rho) \geq 0$ on $\rho > 0$ so long as the metric is asymptotically AdS and the radial NEC is true. The former implies that, at large $\rho$, $A'$ asymptotes to a positive constant. The latter implies that
\begin{equation}
-A''(\rho) \geq 0.\label{necVacA}
\end{equation}
Suppose that $A'$ indeed crosses zero at some point $\rho = \rho_*$. In other words, we are assuming that there exists a $\rho_*$ at which $A'(\rho_*) = 0$ and $A'(\rho) > 0$ if $\rho > \rho_*$. We can Taylor-expand around this point to write
\begin{equation}
A'(\rho) = c_* (\rho - \rho_*)^{p_*} + O\left[(\rho - \rho_*)^{p_* + 1}\right].\label{aPrime}
\end{equation}
$p_* \geq 1$ is the degree of this root of $A'$, and $c_*$ is some real nonzero constant. Now, consider some arbitrarily small number $\varepsilon > 0$. We may approximate $A'$ at $\rho = \rho_* + \varepsilon$ as
\begin{equation}
A'(\rho_* + \varepsilon) \approx c_* \varepsilon^{p_*}.\label{approxAp}
\end{equation}
Our assumption of the positivity of $A'$ implies that $c_*$ must be positive. Furthermore, we may differentiate \eqref{aPrime} to write
\begin{equation}
A''(\rho) = c_* p_* (\rho - \rho_*)^{p_* - 1} + O\left[(\rho - \rho_*)^{p_*}\right].
\end{equation}
Again by considering the point $\rho = \rho_* + \epsilon$, here we may approximate $A''$ as
\begin{equation}
A''(\rho_* + \varepsilon) \approx c_* p_* \varepsilon^{p_* - 1}.\label{approxApp}
\end{equation}
This is certainly positive and thus contradicts the NEC \eqref{necVacA}. Hence, no such $\rho_*$ exists, and $A'(\rho) \geq 0$ on $\rho > 0$.\footnote{Note that this proof actually shows that $A'$ is \textit{strictly} positive $\rho > 0$, although for our purposes it is sufficient to just say that $A'(\rho) \geq 0$.}

\subsection{Thermal Deformations}\label{app:isoTherm}

Now we take the metric \eqref{thermalDomainWall},
\begin{equation}
ds^2 = e^{2A(\rho)}\left[-f(\rho)^2 dt^2 + d\vec{x}^2\right] + d\rho^2,
\end{equation}
where $f$ has a simple root at $\rho = 0$. For this metric to be asymptotically AdS, we must have that, as $\rho \to \infty$,
\begin{equation}
f(\rho) \sim 1,\ \ A(\rho) \sim \frac{\rho}{\ell}.
\end{equation}
This time, we wish to show that
\begin{equation}
\frac{f(\rho)}{A'(\rho)} \geq 0,
\end{equation}
on $\rho > 0$. However, since we are working in the exterior of the black hole, we have that $f(\rho) > 0$. Thus, it is sufficient to prove
\begin{equation}
A'(\rho) \geq 0.
\end{equation}
The radial NEC \eqref{radNecIsoTherm} states that
\begin{equation}
f'(\rho) A'(\rho) - f(\rho)A''(\rho) \geq 0.\label{necTherm}
\end{equation}
We take a similar approach as in the case of vacuum deformations. Given that $A'$ asymptotes to a positive constant, suppose there is a $\rho = \rho_*$ such that $A'(\rho_*) = 0$ and $A'(\rho) > 0$ if $\rho > \rho_*$. We again perform the Taylor expansion \eqref{aPrime}, and we again define a small $\varepsilon > 0$ in the course of showing that the coefficient $c_*$ must be positive. We are also still allowed to differentiate in order to write down the $\rho = \rho_*$ expansion for $A''$.

However, simply writing \eqref{approxApp} does not contradict the NEC. Rather, we must evaluate the left-hand side of \eqref{necTherm} at $\rho = \rho_* + \varepsilon$,
\begin{equation}
f'(\rho_* + \varepsilon)A'(\rho_* + \varepsilon) - f(\rho_* + \varepsilon) A''(\rho_* + \varepsilon) \approx c_* \varepsilon^{p_* - 1}\left[\varepsilon f'(\rho_* + \varepsilon) - f(\rho_* + \varepsilon) p_*\right].
\end{equation}
Because $f(\rho) > 0$ on $\rho > 0$, $f(\rho_* + \varepsilon)$ is approximated by an $O(\varepsilon^0)$ factor. Thus, the second term in the brackets dominates the first. We conclude that this quantity is negative, in contradiction with the radial NEC. This proves that $A'(\rho) \geq 0$ on $\rho > 0$.

\section{Null Energy Condition in the Holographic $p$-Wave Superfluid}\label{app:necpwave}

As we had assumed the radial NEC in our derivation of the monotonic $a$-function, it behooves us to ensure that this is satisfied in the holographic $p$-wave superfluid used as our concrete example.

Consider some null vector $k^\mu$ in \eqref{metricpwave}. Observe that the equation of motion \eqref{eomEYM1} may be used to write its contraction against the matter stress tensor $\tilde{T}_{\mu\nu}$ as
\begin{equation}
k^\mu k^\nu \tilde{T}_{\mu\nu} = \frac{1}{\ell_{\text{P}}^3} k^\mu k^\nu R_{\mu\nu}.
\end{equation}
The radial null vector of \eqref{metricpwave} may be written as
\begin{equation}
k^\mu = \frac{1}{\sigma(r)} \delta^\mu_t + N(r) \delta^\mu_r,
\end{equation}
where the normalization is chosen to avoid factors of $i$. The corresponding contraction reads as
\begin{equation}
k^\mu k^\nu \tilde{T}_{\mu\nu} = \frac{3N(r)^2}{\ell_{\text{P}}^3} \left[\frac{1}{r} \frac{\sigma'(r)}{\sigma(r)} - 2\left(\frac{h'(r)}{h(r)}\right)^2\right].\label{necpwaveRad}
\end{equation}
So the radial NEC is the statement that this expression is nonnegative. Indeed, one of the independent equations of motion \eqref{eompwave2} may be rearranged as
\begin{equation}
\frac{1}{r}\frac{\sigma'}{\sigma} - 2\left(\frac{h'}{h}\right)^2 = \left(\frac{\alpha h^2 w \phi}{\sqrt{3} r N \sigma}\right)^2 + \left(\frac{\alpha h^2 w'}{\sqrt{3}r}\right)^2.
\end{equation}
As all the metric functions are real, the right-hand side of \eqref{necpwaveRad} is manifestly nonnegative by virtue of being a sum of squares.

\section{Numerically Constructing Holographic $p$-Wave Superfluids}\label{app:numerics} 

We briefly discuss the procedure for obtaining the holographic $p$-wave superfluid solutions of \eqref{actEYM}. We assume the metric ansatz \eqref{metricpwave}. We start by expanding the functions around the horizon $r = r_h$:
\begin{align}
m(r) &= m_0 + m_1(r-r_h) + O[(r-r_h)^2],\label{series1}\\
\sigma(r) &= \sigma_0 + \sigma_1(r-r_h) + O[(r-r_h)^2],\\
h(r) &= h_0 + h_1(r-r_h) + O[(r-r_h)^2],\\
\phi(r) &= \phi_0 + \phi_1(r-r_h) + O[(r-r_h)^2],\\
w(r) &= w_0 + w_1(r-r_h) + O[(r-r_h)^2].\label{series5}
\end{align}
Some of these coefficients are already determined by the physics. Specifically, recall that $m$ is defined according to
\begin{equation}
N(r) = -\dfrac{2m(r)}{r^2} + r^2,
\end{equation}
where $N$ is the blackening factor. Since $N(r_h) = 0$, we may write
\begin{equation}
0 = -\frac{2m_0}{r_h^2} + r_h^2 \implies m_0 = \frac{r_h^4}{2}.
\end{equation}
Furthermore, for the one-form potential \eqref{oneform} to be a well-defined one-form at the horizon, we must have that $\left.\mathcal{A}_t\right|_{r = r_h} = 0$ \cite{Kobayashi:2006sb}. Indeed, note that
\begin{equation}
\mathcal{A}^t = g^{tt}\mathcal{A}_t = -\frac{\phi(r)}{N(r)\sigma(r)}
\end{equation}
blows up at $r = r_h$ unless $\mathcal{A}_t$ vanishes there. Thus, we have that $\phi_0 = 0$.

Equipped with these constraints on $\phi_0$ and $m_0$, we plug the series expansions into the equations of motion \eqref{eompwave1}--\eqref{eompwave5} and expand around $r = r_h$. This yields five sets of order-by-order constraints on the coefficients. From the leading $O[(r-r_h)^{-1}]$ terms in \eqref{eompwave3} and \eqref{eompwave5}, respectively,
\begin{equation}
h_1 = 0,\ \ w_1 = 0.\label{derivVanish}
\end{equation}
Applying these, the equations of motion become
\begin{align}
\left(m_1 - \frac{\alpha^2 r_h^3 \phi_1^2}{6\sigma_0}\right) + O(r-r_h) &= 0,\\
\left[\sigma_1 - \frac{\alpha^2 r_h^3 h_0^4 w_0^2 \phi_1^2}{12\sigma_0 (m_1 - 2r_h^3)^2}\right] + O(r-r_h) &= 0,\\
\left[4h_2 + \frac{\alpha^2 r_h^2 h_0^5 w_0^2 \phi_1^2}{12\sigma_0^2 (m_1 - 2r_h^3)^2}\right] + O(r-r_h) &= 0,\\
\left[\frac{3\phi_1}{r_h} + \frac{h_0^4 w_0^2 \phi_1}{2(m_1 - 2r_h^3)} + \frac{\sigma_1 \phi_1}{\sigma_0} + 2\phi_2\right] + O(r-r_h) &= 0,\\
\left[4w_2 + \frac{r_h^4 w_0 \phi_1^2}{4\sigma_0^2 (m_1 - 2r_h^3)^2}\right] + O(r-r_h) &= 0.
\end{align}
$r_h$ and $\alpha$ are ``physical parameters," so we keep them fixed in the course of numerically solving the equations of motion. As for the ``shooting parameters" defining a particular \textit{solution}, the number of such free parameters may be distilled down to the list $\{\sigma_0,h_0,w_0,\phi_1\}$ (as also stated by \cite{Ammon:2009xh}). Higher-order coefficients may be written in terms of this list. For example,
\begin{equation}
\begin{Bmatrix}
m_1 = \dfrac{\alpha^2 r_h^3 \phi_1^2}{6\sigma_0},\qquad\quad
\sigma_1 = \dfrac{3\alpha^2 h_0^4 w_0^2 \sigma_0 \phi_1^2}{r_h^3(12\sigma_0 - \alpha^2 \phi_1^2)^2},\qquad\quad
h_2 = -\dfrac{3\alpha^2 h_0^5 w_0^2 \phi_1^2}{4r_h^4(12\sigma_0 - \alpha^2 \phi_1^2)^2},\vspace{0.25cm}\\
\phi_2 = \dfrac{3\phi_1}{2r_h^3}\left[\dfrac{h_0^4 w_0^2\left(12\sigma_0^2 - \alpha^2 \phi_1^2 (1 + \sigma_0)\right)}{(12\sigma_0 - \alpha^2 \phi_1^2)^2} - r_h^2\right],\qquad
w_2 = -\dfrac{9w_0 \phi_1^2}{4r_h^2 (12\sigma_0 - \alpha^2 \phi_1^2)^2}.
\end{Bmatrix}
\end{equation}
Now we define some small parameters $\varepsilon_{\text{out}},\varepsilon_{\text{in}},\varepsilon_{\text{sing}} > 0$ and a large parameter $r_{\text{bdy}} > 0$. The exterior is taken to be the interval $(r_h + \varepsilon_{\text{out}},r_{\text{bdy}})$, and the interior is $(\varepsilon_{\text{sing}},r_h - \varepsilon_{\text{in}})$. We numerically solve for the functions $\{m,\sigma,h,\phi,w\}$ in each region separately by considering the initial conditions:
\begin{equation}
\text{Exterior}:\ \ \begin{cases}
m(r_h + \varepsilon_{\text{out}}) = m_0 + \varepsilon_{\text{out}} m_1,\\
\sigma(r_h + \varepsilon_{\text{out}}) = \sigma_0 + \varepsilon_{\text{out}} \sigma_1,\\
h(r_h + \varepsilon_{\text{out}}) = h_0,\\
h'(r_h + \varepsilon_{\text{out}}) = 0,\\
\phi(r_h + \varepsilon_{\text{out}}) = \varepsilon_{\text{out}} \phi_1,\\
\phi'(r_h + \varepsilon_{\text{out}}) = \phi_1,\\
w(r_h + \varepsilon_{\text{out}}) = w_0,\\
w'(r_h + \varepsilon_{\text{out}}) = 0.
\end{cases} \ \ \text{Interior}:\ \ \begin{cases}
m(r_h - \varepsilon_{\text{in}}) = m_0 - \varepsilon_{\text{in}} m_1,\\
\sigma(r_h - \varepsilon_{\text{in}}) = \sigma_0 - \varepsilon_{\text{in}} \sigma_1,\\
h(r_h - \varepsilon_{\text{in}}) = h_0,\\
h'(r_h - \varepsilon_{\text{in}}) = 0,\\
\phi(r_h - \varepsilon_{\text{in}}) = - \varepsilon_{\text{in}} \phi_1,\\
\phi'(r_h + \varepsilon_{\text{in}}) = \phi_1,\\
w(r_h - \varepsilon_{\text{in}}) = w_0,\\
w'(r_h - \varepsilon_{\text{in}}) = 0.
\end{cases}\label{initConds}
\end{equation}
Here we have kept the terms in each function's series up to linear order in $\varepsilon$ (or $O(\varepsilon^0)$ for the initial values of derivatives). Recall that $\phi_0 = h_1 = w_1 = 0$. Higher-order terms may be included for more numerical precision, but we find linear order to be sufficient.

If the physical and shooting parameters are specified, then we may construct numerical solutions to the equations of motion. The free parameters are dictated by the physics. For example, \cite{Ammon:2009xh} scans over various mathematically possible values for the free shooting parameters (after fixing the physical parameters) and selects for the ones which exhibit particular asymptotic behavior determined by UV physics.

\subsection{Exterior Solutions}\label{app:ext}

\begin{figure}
\centering
\includegraphics[scale=0.55]{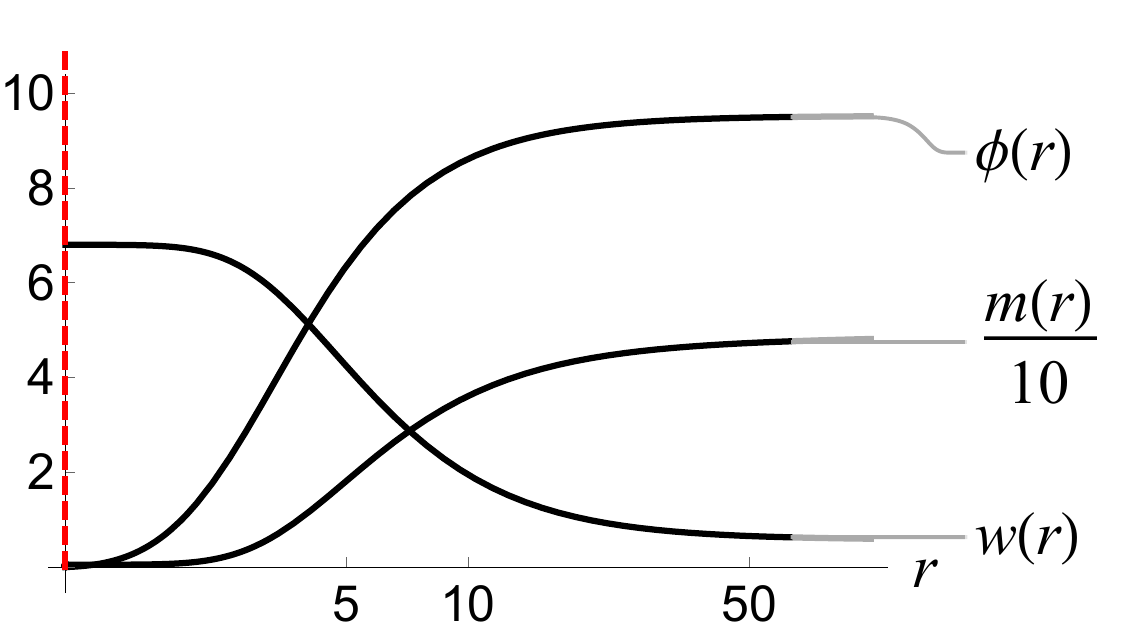}\qquad\includegraphics[scale=0.55]{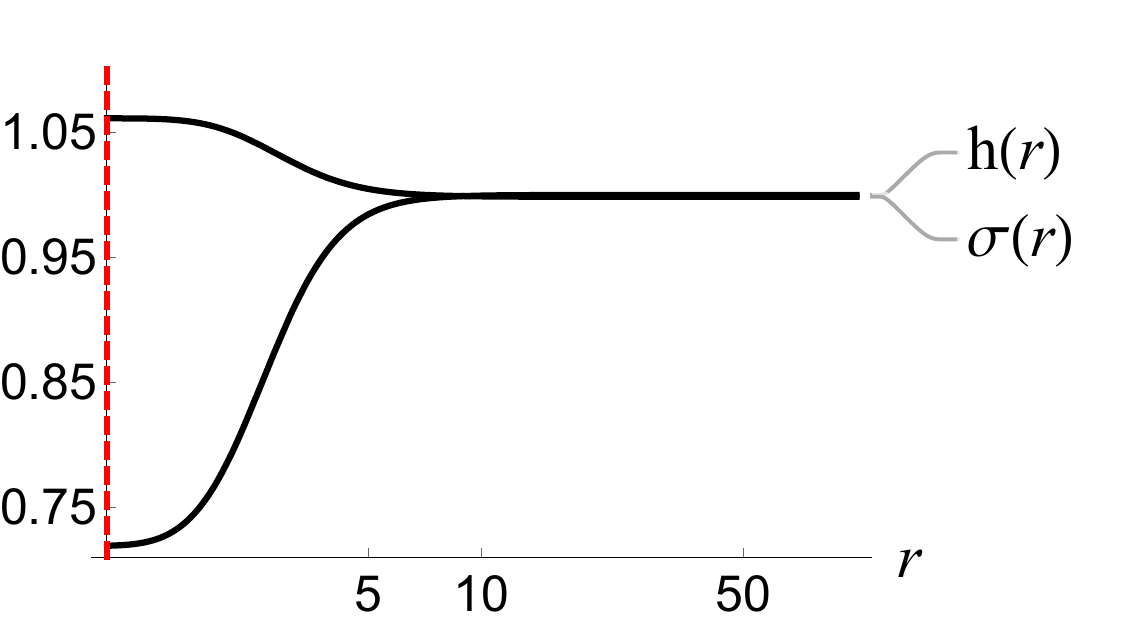}
\caption{The exterior solutions corresponding to the parameters \eqref{paramsShooting}, presented on log plots. The red line represents the horizon. Observe that these are precisely the plots of \cite{Ammon:2009xh}.}
\label{figs:extFuncs}
\end{figure}

As validation, we demonstrate that this procedure reproduces the exterior solution in Figure 1 of \cite{Ammon:2009xh} for the set of parameters:
\begin{equation}
r_h = 1,\ \ \alpha = 0.316,\ \ w_0 = 6.8,\ \ h_0 = 1.061,\ \ \sigma_0 = 0.719,\ \ \phi_1 = 0.131.\label{paramsShooting}
\end{equation}
We show the resulting functions in Figure \ref{figs:extFuncs}.

\subsection{Interior Solutions}\label{app:int}

\begin{figure}
\centering
\includegraphics[scale=0.55]{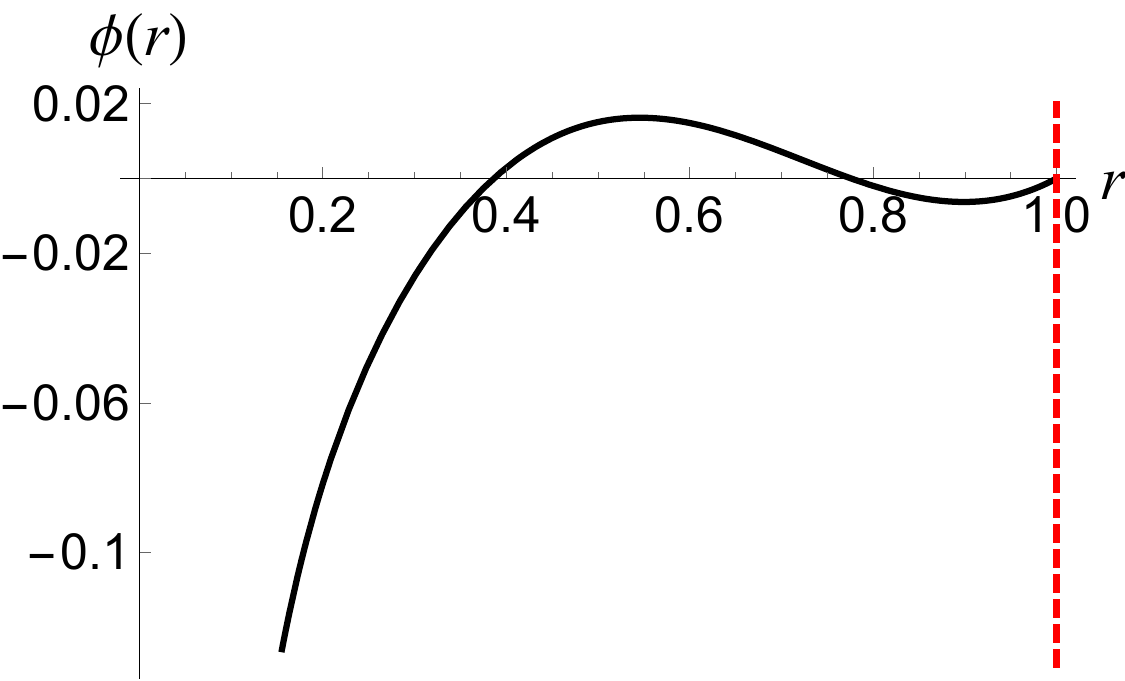}\qquad\includegraphics[scale=0.55]{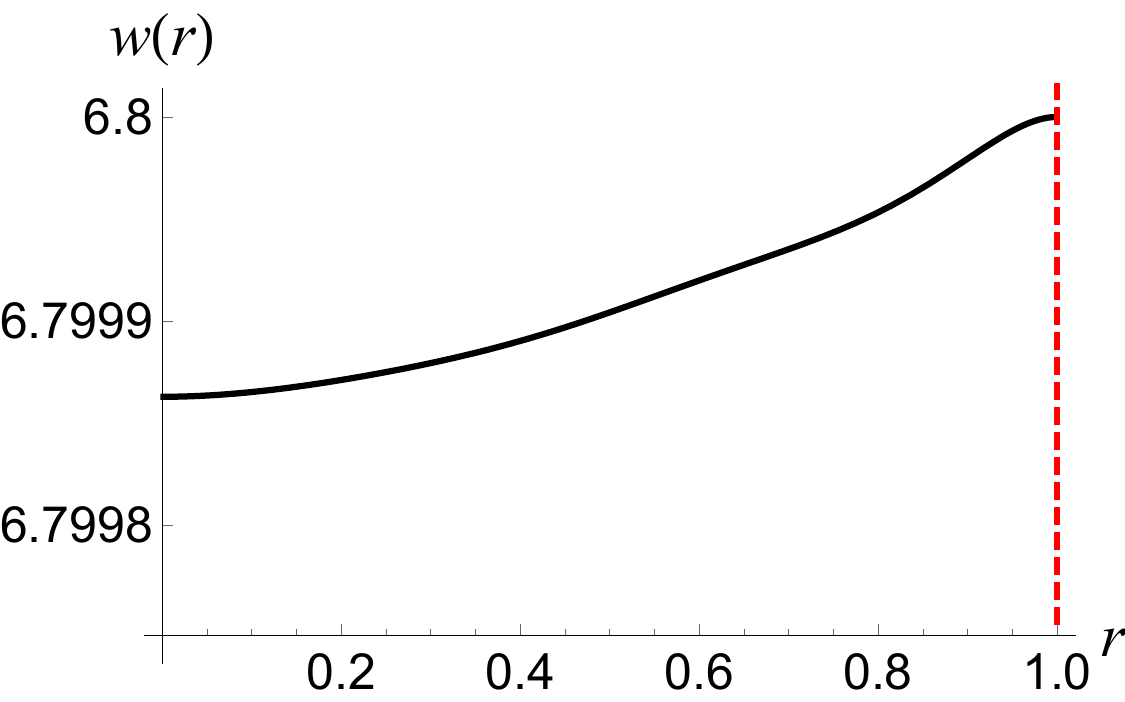}
\caption{The functions describing the one-form potential $\mathcal{A} = \phi(r)\tau^3 dt + w(r) \tau^1 dx$. The $\mathcal{A}_t$ component exhibits some oscillatory behavior before diverging. The $\mathcal{A}_x$ component decreases towards the black hole singularity but numerically appears to asymptote to a finite value.}
\label{figs:interiorPotential}
\end{figure}

We now plot the functions in the interior, a domain which was not included in the analysis of \cite{Ammon:2009xh} but has been the subject of more recent work in a slightly generalized theory \cite{Cai:2021obq}. We start with the functions characterizing the one-form potential: $\phi$ and $w$. These are shown in Figure \ref{figs:interiorPotential}. Observe in particular that $\phi$ exhibits some oscillatory behavior. This is reminiscent of the Josephson oscillators by the charged scalar in the holographic superconductor \cite{Hartnoll:2020fhc,An:2022lvo}.

\begin{figure}
\centering
\includegraphics[scale=0.55]{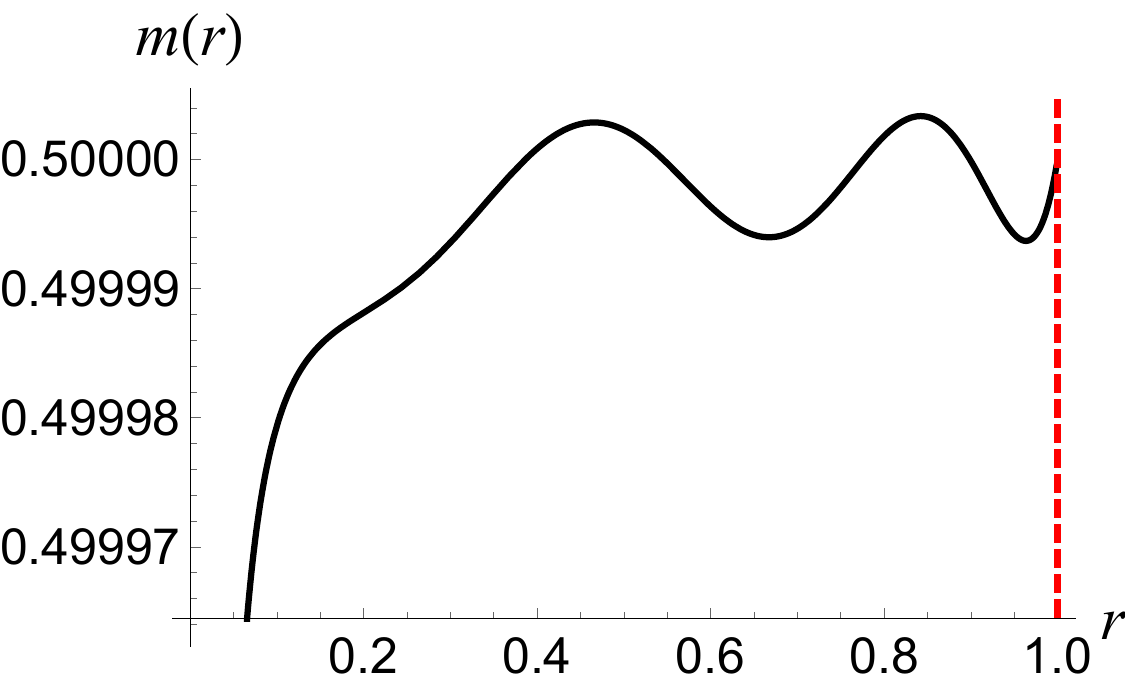}\qquad\includegraphics[scale=0.55]{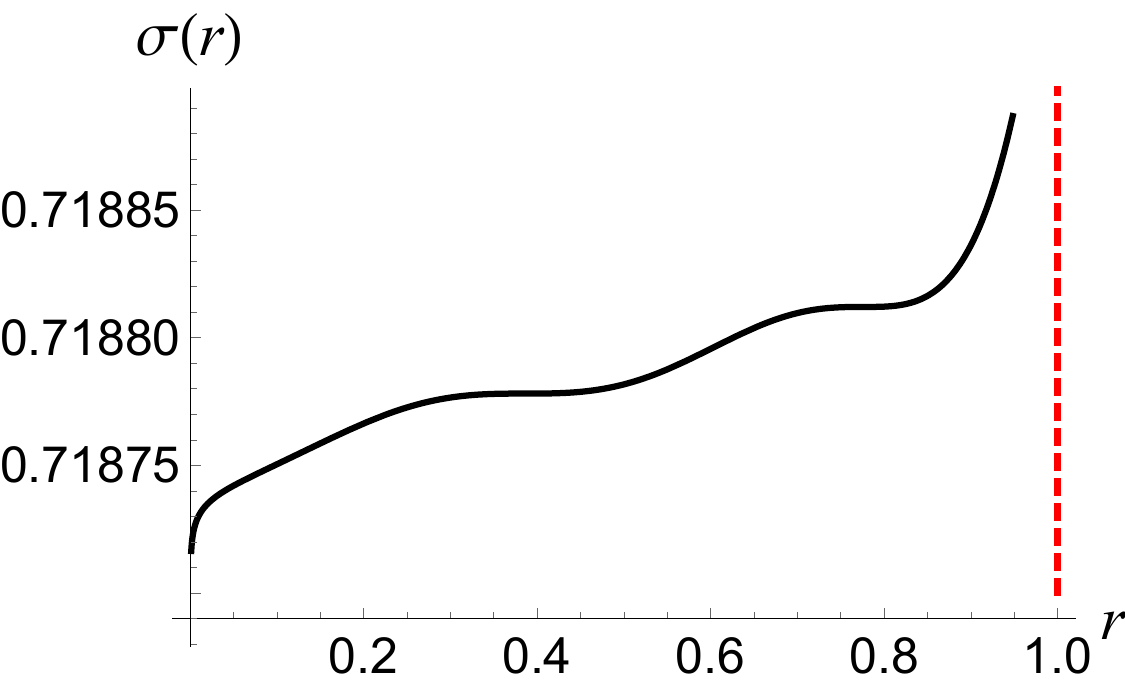}
\caption{The metric functions $m$ and $\sigma$ which reflect the oscillatory behavior of $\phi$ in the black hole interior.}
\label{figs:interiorMetOscillations}
\end{figure}

The oscillatory behavior imprints upon the metric, as well. This is evident in plots of $m$ (corresponding to the blackening function $N$) and $\sigma$ (describing the $a$-function)---see Figure \ref{figs:interiorMetOscillations}.

\begin{figure}
\centering
\includegraphics[scale=0.55]{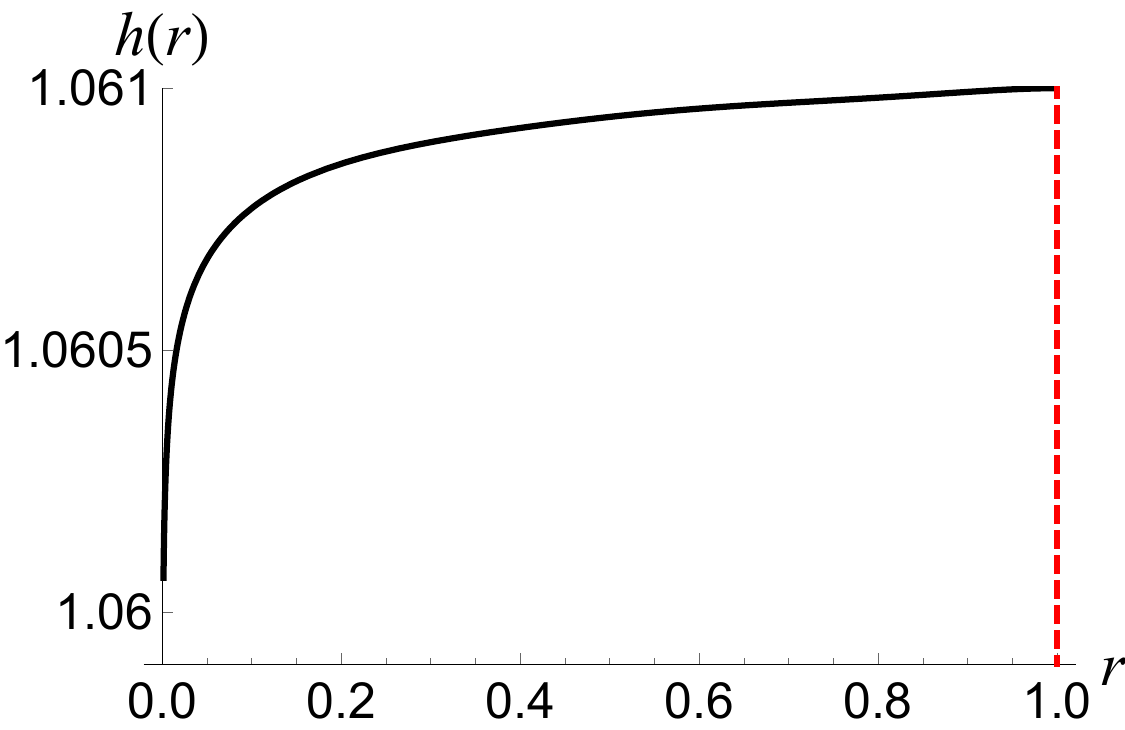}\qquad\includegraphics[scale=0.55]{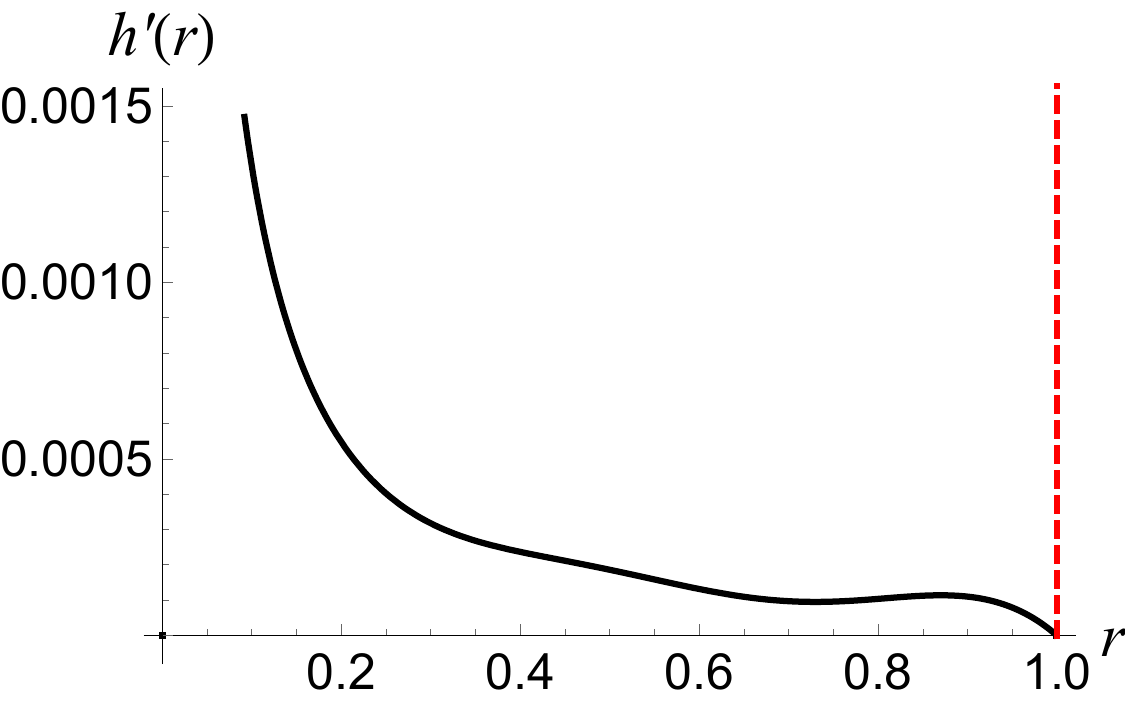}
\caption{The metric function $h$ describing anisotropy in the black hole interior and its derivative $h'$. Note that $h$ monotonically decreases towards as we flow towards the singularity. Furthermore, $h'$ appears to rise in an unbounded manner, corresponding to an increasing gradient of $h$.}
\label{figs:interiorAniso}
\end{figure}

Lastly, we may ask what happens to the anisotropy, which is described by $h$. We plot both $h$ and its derivative in Figure \ref{figs:interiorAniso}. While the solution appears to become more isotropic ($h \to 1$) as we approach the singularity, it is not clear whether $h$ settles at $1$ or goes below $1$. The steepness seen in the plot of $h'$ would suggest the latter, but this may also be an artifact of our numerical approach. More refined analysis (numerical or otherwise) would be needed to answer this question.

\end{appendices}
\vfill
\bibliographystyle{jhep}
\bibliography{references.bib}
\end{document}